\documentclass[aps,twocolumn,prb,reprint,floatfix,showpacs,lengthcheck,citeautoscript,superscriptaddress,nofootinbib]{revtex4-2}
\usepackage{amsmath}
\usepackage{amssymb}
\usepackage{graphics}
\usepackage{graphicx}
\usepackage{bm}
\usepackage{float}
\usepackage{xcolor}
\usepackage{times}
\usepackage{notes2bib}
\usepackage{verbatim}
\usepackage[none]{hyphenat}
\usepackage{epstopdf}
\usepackage{tikz}
\usepackage{braket}
\usepackage{bm}
\usepackage[colorlinks=true,
linkcolor=blue, 
citecolor=blue, 
urlcolor=blue,
bookmarks=true,
hypertexnames=true]{hyperref}

\newcommand{\af}{\alpha}
\newcommand{\ve}{\varepsilon}

\newcommand{\ci}{\textsf{i}}
\newcommand{\mi}{\textsf{m}}
\newcommand{\li}{\textsf{l}}

\newcommand{\bigdot}{\boldsymbol{\cdot}}

\newcommand{\sq}{\sqrt{3}}

\newcommand{\hw}{\hbar\omega}

\newcommand{\moi}{\leqslant}
\newcommand{\dk}{\frac{d\bm{k}}{(2\pi)^3}}

\newcommand{\e}{\epsilon}
\newcommand{\w}{\omega}

\newcommand{\ti}{\theta}
\newcommand{\p}{\partial}
\newcommand{\scp}{\scriptscriptstyle}
\newcommand{\chil}{\chi_{ij}^{\scriptscriptstyle (1)}}
\newcommand{\chiq}{\chi_{ijk}^{\scriptscriptstyle (2)}(-2\omega;\omega,\omega)}
\newcommand{\chiqs}{ \chi_{ijk}^{\scriptscriptstyle (2)} }
\newcommand{\xyz}{\chi_{xyz}^{\scriptscriptstyle (2)}}

\newcommand{\tw}{\tilde{\omega}}
\newcommand{\Tr}{\text{Tr}}
\newcommand{\bx}{\bm{e}_x}
\newcommand{\by}{\bm{e}_y}
\newcommand{\bz}{\bm{e}_z}
\newcommand{\fce}{\textsf{ce}}
\newcommand{\fir}{\textsf{ir}}
\newcommand{\fim}{\textsf{im}}
\newcommand{\vphi}{\varphi}

\newcommand{\dv}{\int\hspace{-1.5mm}d\bm{r}\,} 

\begin{document}
    \title{Perturbative second-order optical susceptibility of bulk materials:\\a symmetry-enforced return to non-orthogonal localized basis sets}
	\author{Angiolo Huamán}
    \email{ah200@uark.edu}
\affiliation{Department of Physics, University of Arkansas, Fayetteville, Arkansas 72701, USA and\\
	MonArk NSF Quantum Foundry, University of Arkansas, Fayetteville, Arkansas 72701, USA}
    \author{Luis Enrique Rosas-Hernandez}
\affiliation{Department of Physics, University of Arkansas, Fayetteville, Arkansas 72701, USA and\\
	MonArk NSF Quantum Foundry, University of Arkansas, Fayetteville, Arkansas 72701, USA}
	\author{Salvador Barraza-Lopez}
    \email{sbarraza@uark.edu}
\affiliation{Department of Physics, University of Arkansas, Fayetteville, Arkansas 72701, USA and\\
	MonArk NSF Quantum Foundry, University of Arkansas, Fayetteville, Arkansas 72701, USA}
\affiliation{Institute for Solid State Physics, University of Tokyo, 5-1-5 Kashiwanoha, Kashiwa, Chiba 277-8581, Japan}
	\begin{abstract}

The second-order optical susceptibility of semiconductors $\chiq$ finds application in metrology, spectroscopy, telecommunications, material characterization, and quantum information. Pioneering calculations of $\chiq$ utilized non-orthogonal Gaussian orbitals centered at atoms. That formulation transitioned into plane-wave-based algorithms as time went by. As of late, nevertheless, multiple tools for calculating optical susceptibilities have recast the problem using Wannier ({\em i.e.}, {\em localized}) orbitals, making a comeback onto frameworks based on localized basis sets. Here, we present an approach for calculating $\chiq$ reliant on numerical pseudoatomic orbitals (PAOs) within perturbation theory in the velocity gauge. Its salient feature is a calculation of `Slater-Koster-like' two-center integrals of the momentum operator in between PAOs identified by symmetry. The approach was successfully tested on paradigmatic cubic silicon carbide (3C-SiC) and gallium arsenide, for which linear responses are contributed as well.

	\end{abstract}
	\date{\today}
	\maketitle
\section{Introduction}
Second harmonic generation (SHG)~\cite{Franken} arises from the optical second-order susceptibility tensor $\chiq$. It is relied upon to change frequency from laser sources of light, and it has scientific and technological relevance for metrology, spectroscopy and telecommunications, for structural~\cite{Yao2021} and magnetic~\cite{Jeong2024} material characterization, and for generating entangled photon pairs~\cite{Hillery} for quantum information applications by means of parametric down conversion~\cite{Kwiat1995,Kwiat2001}.

Theoretical {\em ab initio} formalisms for computing $\chiq$ in semiconductors generally follow two main routes: (i) perturbation theory within the independent-particle approximation (IPA), formulated in either the length or velocity gauge~\cite{Sipe1993,Hughes1996}, and with the possibility of intraband transitions~\cite{Aversa1995} and electro-optical effects~\cite{Sipe2000}; or (ii) time-dependent density functional theory (TDDFT)~\cite{Luppi2010,Tancogne-Dejean2016}.

Most methodologies for calculating the optical second-order susceptibility based on perturbation theory rely on orthogonal basis sets, typically plane waves or post-processed Wannier functions. Examples include the Optic utility and the TINIBA code~\cite{Cabellos2009} for ABINIT~\cite{Gonze2002}, or the YAMBO code~\cite{Gruning2014,Marini2009}, which interfaces with QUANTUM ESPRESSO~\cite{Giannozzi_2017}, ABINIT, or CASTEP~\cite{Milman2010}. A recent methodology for SHG based on localized basis sets is provided in Ref.~\cite{Li2022}.

Early implementations based on the $2n+1$ theorem within density-functional perturbation theory used a Wannier representation obtained through direct total-energy minimization to evaluate the static limit of $\chiq$~\cite{DalCorso1994}. This framework was later extended in Ref.~\cite{DalCorso1996}, which combined the $2n+1$ formalism with TDDFT to compute $\chiq$ for several zinc-blende semiconductors. Subsequent real-time TDDFT developments enabled the computation of $\chiq$ directly from the time evolution of the polarization. For example, the authors of Ref.~\cite{Takimoto2007} used real-time TDDFT with pseudo-atomic orbitals (PAOs) to evaluate $\chiq$, while the 2Light code~\cite{Luppi2010}—interfaced with ABINIT—can compute macroscopic nonlinear susceptibilities within a many-body TDDFT framework. The Octopus package also offers a TDDFT implementation reliant on linear combinations of PAOs as trial functions~\cite{Tancogne-Dejean2020}. YAMBO, in turn, is a dedicated tool for many-body calculations, incorporating TDDFT and solving the Bethe$-$Salpeter equation. TDDFT-based approaches for high-harmonic generation using Gaussian orbitals have been explored as well; see Refs.~\cite{Luppi2022} and~\cite{Yasin2021}.

The SIESTA DFT package~\cite{Soler2002,Garcia2020} currently contains {\em linear} optical response capabilities~\cite{Junquera2001}. Here we present a formalism for the calculation of the second-order optical susceptibility $\chiq$ within the IPA and perturbation theory relying on a non-orthogonal PAO basis set~\cite{Junquera2001,Ozaki1,Ozaki2}, which we call {\it Nonlinear Algorithms by Post Processing in SIESTA}. This framework is particularly well suited for simulations requiring significant vacuum—such as molecules, nanotubes, surfaces, or 2D materials—under the assumption that continuum states are not activated, and it does not require Wannierization postprocessing.

Excitonic effects arising from electron$-$hole interactions generated when an electron is promoted from the valence to the conduction bands~\cite{Hanke1980,Berkelbach2013} can be incorporated through several methodologies: by solving the GW-Bethe-Salpeter equations, as it was done in MoS$_2$~\cite{Trolle2014,Mkrtchian2019} and NbOI$_2$~\cite{Xuan2024}; through Hartree$-$Fock corrections applied to GW single-particle levels~\cite{Gruning2014}; or by many-body wavefunction expansions in terms of single electron$-$hole pair excitations~\cite{Chang2001}. These many body effects entail the appearance of features below the optical absorption edge~\cite{KittelQTS,Ruan2024}. The IPA-based formalism showcased here represents a first step toward an eventual incorporation of many-body effects (see, {\em e.g.}, Ref.~\cite{PhysRevB.109.155437}), and it can already find use in symmetry analysis of SHG responses of 2D materials \cite{Hiroyuki2025}.

The paper is organized as follows: An explicit procedure to construct Bloch eigenstates from a non-orthogonal PAO basis set is presented in section~\ref{nonorto}. Section~\ref{formalism} contains the equations describing the second-order optical susceptibility reliant on a non-orthogonal basis set, and symmetry-based methods for calculating momentum matrix elements of the Bloch eigenstates and two-center integrals between PAOs are outlined in section~\ref{two_center_intg}. Section~\ref{sec:FBZ_intg} gives details on the integration of $\chiq$ within a time-reversal-symmetric irreducible first Brillouin zone. Computational details and the algorithm's flow are included in section~\ref{sec:comp_details}. Results on zinc-blende silicon carbide and gallium arsenide, as well as additional benchmarks concerning linear optical responses are presented in Section~\ref{results}. Section~\ref{conclusions} offers conclusions.

Appendix~\ref{chi-derivation} contains a detailed account on how to obtain the equations for the linear and second-order susceptibility tensors, and Appendix~\ref{Dmatrices} summarizes the rotation matrices for spherical harmonics used within the main text.

One main difficulty for non-experts in this field is the multiple use of similar symbols within different contexts (for example, $i$ representing the imaginary unit or a Cartesian direction, $k$ the magnitude of crystal momentum or a Cartesian direction, $l$ representing total angular momentum of a cartesian direction, and so on). Therefore, and risking alienating versed readers, a deliberate effort to use different fonts to tell the different context in which a given label is being used  was undertaken. We hope such effort results in an easier flow.

\begin{table*}[tb]
	\caption{Real spherical harmonics with the Condon-Shortley phase and sum convention given in equation \eqref{eq:RealHarmonics}~\cite{MartinES} up to $d-$orbitals. Roman numbers at the far left indicate the relative placement of orbitals in Hamiltonian and overlap matrices, and in other transformation matrices.\label{ta:SphericalHarmonics}}
		\begin{tabular}{c|c}
			\hline
			\hline
			1) $s(\theta,\phi)\equiv Y_{00}(\theta,\phi)=\frac{1}{2\sqrt{\pi}}$ &
2) $d_{yz}(\theta,\phi)\equiv s^-_{21}(\theta,\phi)=-\sqrt{\frac{15}{4\pi}}\sin\theta\sin\phi\cos\theta=-\sqrt{\frac{15}{4\pi}}\frac{yz}{r^2}$ \\
			1) $p_y(\theta,\phi)\equiv s^-_{11}(\theta,\phi)=-\sqrt{\frac{3}{4\pi}}\sin\theta\sin\phi=-\sqrt{\frac{3}{4\pi}}\frac{y}{r}$ &
3) $d_{3z^2-r^2}(\theta,\phi)\equiv Y_{20}(\theta,\phi)= \sqrt{\frac{5}{16\pi}} \left(3\cos^2\theta -1\right)=\sqrt{\frac{5}{16\pi}}\frac{3z^2-r^2}{r^2}$ \\
			2) $p_z(\theta,\phi)\equiv Y_{10}(\theta,\phi)=\sqrt{\frac{3}{{4\pi}}}\cos\theta=\frac{3}{\sqrt{4\pi}}\frac{z}{r}$ &
4) $d_{xz}(\theta,\phi)\equiv s^+_{21}(\theta,\phi)=-\sqrt{\frac{15}{4\pi}}\sin\theta\cos\phi\cos\theta=-\sqrt{\frac{15}{4\pi}}\frac{xz}{r^2}$\\
            3) $p_x(\theta,\phi)\equiv s^+_{11}(\theta,\phi)=-\sqrt{\frac{3}{4\pi}}\sin\theta\cos\phi=-\sqrt{\frac{3}{4\pi}}\frac{x}{r}$ &
5) $d_{x^2-y^2}(\theta,\phi)\equiv s^+_{22}(\theta,\phi)=\sqrt{\frac{15}{16\pi}}\sin^2\theta\left(\cos^2\phi- \sin^2\phi\right)$\\
			1) $d_{xy}(\theta,\phi)\equiv s^-_{22}(\theta,\phi)=\sqrt{\frac{15}{4\pi}}\sin\theta\cos\phi \sin\theta\sin\phi= \sqrt{\frac{15}{4\pi}}\frac{xy}{r^2}$ &
$= \sqrt{\frac{15}{16\pi}}\frac{x^2-y^2}{r^2}$\\
			\hline
			\hline
		\end{tabular}
\end{table*}

\section{Bloch eigenstates from non-orthogonal pseudo-atomic orbitals}\label{nonorto}
The process for building Bloch wavefunctions $\psi^0_{n\bm{k}}(\bm{r})$ from non-orthogonal PAOs is presented now. This approach is currently restricted to spinless systems (for which time-reversal symmetry holds), and it involves these steps: (1) retrieving the PAOs' radial functions from SIESTA's output, (2) reconstructing the PAOs in a real space grid from their radial and angular dependencies, and (3) extracting the Hamiltonian and overlap matrix elements from SIESTA's output, from which unperturbed eigenvalues and eigenvectors at desired $k-$points are obtained).

To construct the real-space representation of the PAOs, consider a unit cell containing \textsf{nce} chemical species. Though the process is generic, it will be exemplified with 3C-SiC in what follows. The basis size for each chemical species  \textsf{ce} (ranging from 1 to  \textsf{nce}) depends on two factors: the \verb|PAO.BasisSize| flag (which can take labels  \verb|SZ|, \verb|SZP|, \verb|DZ|, \verb|DZP|, \verb|TZ|, \verb|TZP|, and so on \cite{Junquera2001}), and its electronic valence configuration. Setting the label \textsf{ce}=1 to refer to silicon and \textsf{ce}=2 to label carbon in what follows, the valence configuration for the first element is $3s^23p^2$, which makes the polarization orbitals (if they are added by including the letter \verb|P| in the basis specification) have a $d-$character. Similar to silicon, the electronic valence configuration for carbon ($2s^22p^2$) features $s$ and $p$ orbitals, and its polarization orbitals will include an extra angular momentum channel ($d$), too. The angular momentum of polarization orbitals may be different for other chemical elements.

The letter \verb|S|, \verb|D|, or \verb|T| at the start of the basis specification indicates that the number of radial functions in the valence channels is \textsf{nrv}=1, 2, or 3, respectively (\textsf{ir}=1,...,\textsf{nrv}; the number of radial functions for the polarization orbital is set to \textsf{ir}=\textsf{nrp}=1 when using standard basis flags in SIESTA). For SiC, orbitals with total angular momentum \textsf{l}=0 or 1 belong to the valence, and orbitals with \textsf{l}=2 are polarization ones.

\begin{figure}[tb]
	\centering
\includegraphics[width=\columnwidth]{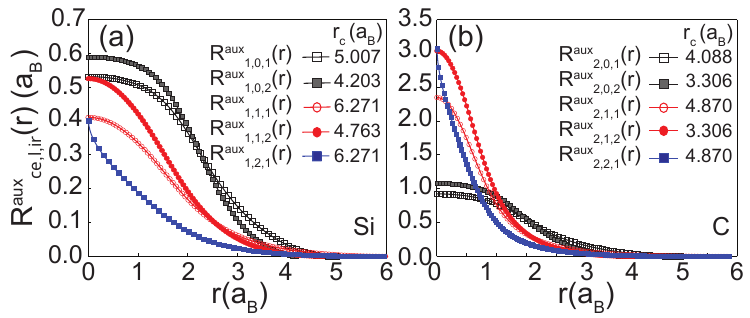}
	\caption{Typical radial functions: (a) $R^{aux}_{1,\textsf{l}(1),\textsf{ir}(\textsf{l},1)}(r)$ for Si; $a_B=0.529177$ \AA{} is the Bohr radius. (b) $R^{aux}_{2,\textsf{l}(2),\textsf{ir}(\textsf{l},2)}(r)$ for C. Cutoff radii are written, too.\label{radial}}
\end{figure}

SIESTA's output files with the \verb|ion| extension (one per chemical species \textsf{ce}) contain ``auxiliary'' radial functions up to length cutoffs $r_c(\textsf{ce},\textsf{l}(\textsf{ce}),\textsf{ir}(\textsf{l},\textsf{ce}))$, one per radial function, over a one-dimensional grid containing 500 points. For the \verb|DZP| flag, both silicon and carbon have two radial functions for both the $s$ (\textsf{l}=0) and $p$ (\textsf{l}=1) channels, and one radial function for the d (\textsf{l}=2) channel. To avoid an abrupt truncation of the radial function at $r_c(\textsf{ce},\textsf{l}(\textsf{ce}),\textsf{ir}(\textsf{l},\textsf{ce}))$, soft-confinement settings in SIESTA were configured to \verb|PAO.SoftDefault| \verb|.true.| and \verb|PAO.SoftInnerRadius 0.75|. We call these $R^{aux}_{\textsf{ce},\textsf{l}(\textsf{ce}),\textsf{ir}(\textsf{l},\textsf{ce})}(r)$ and show them in figure \ref{radial} for silicon and carbon. The radial functions are:
\begin{equation}\label{eq:full_Rar}
R_{\textsf{ce},\textsf{l}(\textsf{ce}),\textsf{ir}(\textsf{l},\textsf{ce})}(r)=r^{\textsf{l}}R^{aux}_{\textsf{ce},\textsf{l}(\textsf{ce}),\textsf{ir}(\textsf{l},\textsf{ce})}(r).
\end{equation}

The angular dependence of PAOs is given by linear combinations of spherical harmonics $Y_{\textsf{l}\textsf{m}}(\ti,\phi)$ defined with the Condon-Shortley phase convention ($\textsf{m}$ is the projection of the angular momentum, and ranges between $-\textsf{l}$ and $\textsf{l}$):
\begin{equation*}
	Y_{\textsf{l}\textsf{m}}(\theta,\phi)\equiv (-1)^\textsf{m}\sqrt{\frac{2\textsf{l}+1}{4\pi}\frac{(\textsf{l}-\textsf{m})!}{(\textsf{l}+\textsf{m})!}}P_{\textsf{l}}^\textsf{m}(\cos\theta)\text{e}^{\textsf{i}\textsf{m}\phi},
\end{equation*}
where $\theta$ and $\phi$ are the polar and azimuthal angles, respectively, and $P_{\textsf{l}}^\textsf{m}(\cos\theta)$ is an associated Legendre function. The letter $\textsf{i}$ (written in `sans' font) is reserved for the imaginary unit in this manuscript. The $(-1)^\textsf{m}$ phase is absent in classical descriptions of spherical harmonics \cite{JacksonCE}, but it is useful when dealing with quantum mechanical representations of angular momentum. The standard linear combinations can be found in Ref.~\cite{MartinES}:
\begin{align}\label{eq:RealHarmonics}
	s^-_{\textsf{l}\textsf{m}}(\theta,\phi) &=\frac{Y_{\textsf{l}\textsf{m}}(\theta,\phi)-(Y_{\textsf{l}\textsf{m}}(\theta,\phi))^*}{\sqrt{2}i}, \text{ and}\\
	s^+_{\textsf{l}\textsf{m}}(\theta,\phi) &=\frac{Y_{\textsf{l}\textsf{m}}(\theta,\phi)+(Y_{\textsf{l}\textsf{m}}(\theta,\phi))^*}{\sqrt{2}}.\nonumber
\end{align}
Equation \eqref{eq:RealHarmonics} with the Condon-Shortley spherical harmonics listed in Ref.~\cite{VarshalovichQTAM} gives the results in Table \ref{ta:SphericalHarmonics}, where the standard $s$, $p$ and $d$ notations are used.

PAOs centered at the origin of coordinates are products of the radial functions $R_{\textsf{ce},\textsf{l}(\textsf{ce}),\textsf{ir}(\textsf{l},\textsf{ce})}(r)$ in equation~\eqref{eq:full_Rar} and the angular functions listed in Table \ref{ta:SphericalHarmonics}, which contain canceling powers of $r$ for all values of \textsf{l}. This way, explicit PAOs  are products of the spherical functions in Table \ref{ta:SphericalHarmonics} without their $1/r^\li$ factors and the ``auxiliary'' radial functions $R^{aux}_{\textsf{ce},\textsf{l}(\textsf{ce}),\textsf{ir}(\textsf{l},\textsf{ce})}(r)$ from the \verb|ion| files. Every PAO $\varphi_{\textsf{ce},\textsf{l},\textsf{ir},\textsf{im}}(\bm{r})$ listed in Table \ref{ta:PAOs} is thus uniquely characterized by four integers that indicate the chemical species, total angular momentum, radial function number, and a sequential index \textsf{im} given  for each value of \textsf{l} in Table \ref{ta:SphericalHarmonics}. These PAOs are calculated in a 3D real-space  grid of size $\textsf{ngp}^3$ for numerical integration later on; they are normalized such that their square integrates to 1.

\begin{figure}[tb]
	\centering
\includegraphics[width=\columnwidth]{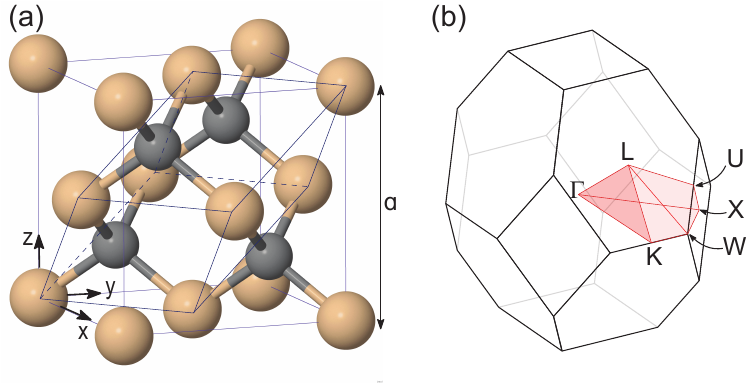}
	\caption{(a) 3C-SiC conventional (cubic) unit  cell (with lattice parameter $a$ shown) and primitive unit cell (parallelepiped). (b) FBZ with high-symmetry points indicated. The irreducible first Brillouin zone including time reversal symmetry is highlighted, too. \label{fig1}}
\end{figure}

The methodology we developed will be tested in 3C-SiC, whose primitive unit cell and first Brillouin zone (FBZ) are shown in figure \ref{fig1}. The basis atoms at locations $\bm{\delta}_{\textsf{ia}(\textsf{ce})}$ (with \textsf{ia} ranging from 1 up to the number of atoms for a given chemical species \textsf{ce} in the unit cell) are read from SIESTA and their corresponding PAOs are displaced by $\bm{\delta}_{\textsf{ia}(\textsf{ce})}$ within the numerical mesh accordingly. For the materials here considered, there is one atom per chemical species, so $\boldsymbol{\delta}_{\textsf{ce}}$ is enough to characterize the atomic locations.

The granularity in the description of PAOs will be used explicitly when addressing two-center integrals of the momentum operator in section \ref{two_center_intg}. For the time being, an equivalent variable $\af\equiv(\fce,\li,\fir,\fim)$ running from 1 to $\mathcal{N}$ encompassing all orbitals within the unit cell helps unburden notation. The variable $\alpha$ orders orbitals per chemical species and angular momentum, and consecutively locates orbitals for a given value of \textsf{l} and \textsf{im} for each of the radial functions available for these two variables.

Upon completing a self-consistent electronic structure calculation, SIESTA also generates a NetCDF-format output file with extension \verb|HSX|. Using the \verb|hsx2hs| utility provided by SIESTA, the \verb|HSX| file is converted into a plain text file with extension \verb|HS|, containing the Hamiltonian and overlap matrix elements between pseudo-orbitals as well as atomic and lattice information necessary to construct tight-binding Hamiltonian  and overlap matrices, $H_0(\bm{k})$ and $S_0(\bm{k})$. Eigenenergies and eigenvectors are obtained by solving the generalized eigenvalue problem:
\begin{equation}\label{eq:GEP}
	H_0(\bm{k}) \bm{C}^0_{n}({\bm{k}}) = \epsilon^0_n(\bm{k}) S_0(\bm{k}) \bm{C}^0_{n}({\bm{k}}),
\end{equation}
where $\epsilon_n^0 (\bm{k})=\hbar \omega_n^0(\bm{k})$, $n$ is a  band index ($1\leq n\leq \textsf{nbands}$) and \textsf{nbands} is the number of bands being considered in the calculation, a number smaller or equal to the total number of numerical orbitals $\mathcal{N}$ utilized on a given calculation, and larger than $\textsf{nvb}$, the number of valence bands. The choice of \textsf{nbands} depends on the energy range where the optical response will be calculated. $\bm{k}$ is a $k$-point within the first Brillouin zone (FBZ), which can be tuned by the user with the \textsf{nk} input flag. The eigenvector $\bm{C}_n^0(\bm{k})$ contains the coefficients to build Bloch eigenstates out of PAOs.

\begin{table*}[tb]
\caption{PAOs for SiC with a DZP basis set. \textsf{ce}=1 (Si) or 2 (C). The index \textsf{ir} takes values 1 and 2 for valence electrons ($s$ and $p$ for this material) when using this basis set. Numbers on parentheses at the right are the number of orbitals, which add up to $\mathcal{N}=26$.\label{ta:PAOs} }
		\begin{tabular}{c|c|c}
			\hline
			\hline
$\varphi_{\fce,0,ir,1}(\bm{r})\equiv  \frac{1}{2\sqrt{\pi}}R^{aux}_{\textsf{ce},0,ir}(r)$ (4) &
$\varphi_{\fce,1,ir,3}(\bm{r})\equiv -\sqrt{\frac{3}{4\pi}}xR^{aux}_{\textsf{ce},1,ir}(r)$ (4)&
$\varphi_{\fce,2,1,3}(\bm{r})\equiv  \sqrt{\frac{5}{16\pi}}(3z^2-r^2)R^{aux}_{\textsf{ce},2,1}(r)$ (2) \\
$\varphi_{\fce,1,ir,1}(\bm{r})\equiv -\sqrt{\frac{3}{4\pi}}yR^{aux}_{\textsf{ce},1,ir}(r)$ (4)&
$\varphi_{\fce,2,1,1}(\bm{r})\equiv  \sqrt{\frac{15}{4\pi}}xyR^{aux}_{\textsf{ce},2,1}(r)$ (2)&
$\varphi_{\fce,2,1,4}(\bm{r})\equiv -\sqrt{\frac{15}{4\pi}}xzR^{aux}_{\textsf{ce},2,1}(r)$ (2)\\
$\varphi_{\fce,1,ir,2}(\bm{r})\equiv  \sqrt{\frac{3}{4\pi}}zR^{aux}_{\textsf{ce},1,ir}(r)$ (4)&
$\varphi_{\fce,2,1,2}(\bm{r})\equiv -\sqrt{\frac{15}{4\pi}}yzR^{aux}_{\textsf{ce},2,1}(r)$ (2)&
$\varphi_{\fce,2,1,5}(\bm{r})\equiv  \sqrt{\frac{15}{16\pi}}(x^2-y^2)R^{aux}_{\textsf{ce},2,1}(r)$ (2)\\
			\hline
			\hline
		\end{tabular}
\end{table*}

Sizes of matrices $H_0(\bm{k})$ and $S_0(\bm{k})$ are $\mathcal{N}\times \mathcal{N}=26\times 26$ when a double$-\zeta$ plus polarization (\verb|DZP|) basis is used on the SiC unit cell (see Table \ref{ta:PAOs}). The orbital basis size  $\mathcal{N}$ decreases to 18 or increases to 34 when \verb|SZP| or \verb|TZP| basis are employed, respectively. For comparison, a cuttof of $300$, $600$, or $900$ eV in plane-wave codes presupposes a basis set containing $\mathcal{N}=260$, $\mathcal{N}=690$, or $\mathcal{N}=1264$ states, respectively, which is between one and two orders of magnitude larger (Wannierization approaches are used to reduce the basis size).

 The  Bloch sum is then constructed as:
\begin{eqnarray}\label{eq:lcao}
\varphi_{\alpha\bm{k}}(\bm{r}) = \frac{1}{\sqrt{N_n}} \sum_{n_1}\sum_{n_2}\sum_{n_3}  e^{\textsf{i}\bm{k}\cdot \bm{R}_{n_1,n_2,n_3}} \times\nonumber\\
\varphi_\alpha(\bm{r}-\bm{R}_{n_1,n_2,n_3}-\boldsymbol{\delta}_\alpha),
\end{eqnarray}
where $\bm{\delta}_{\alpha}=\bm{\delta}_{\textsf 1}$ when $1\moi\alpha\moi 13$, and $\bm{\delta}_{\alpha}=\bm{\delta}_{\textsf 2}$ when $14\moi\alpha\moi 26$ for this material and basis size. $\bm{R}_{n_1,n_2,n_3}\equiv n_1\bm{a}_1+n_2\bm{a}_2+n_3\bm{a}_3$, with $\bm{a}_j$ ($j=1,2,3$) primitive lattice vectors. The indexes $n_j$ ($j=1,2,3$) are defined such that $-n_{j}^{\prime} \leq n_j \leq n_{j}^{\prime}$ (the $n'_j$ relate to the size of the sample). The $N_n$ is the standard normalization factor. With the Bloch sums defined above, the unperturbed Bloch wavefunction therefore reads:
\begin{equation}\label{kwf2}
	\psi^0_{n\bm{k}}(\bm{r})=\sum_{\alpha=1}^{\mathcal{N}} C^{0}_{\alpha n}(\bm{k}) \varphi_{\alpha\bm{k}}(\bm{r}).
\end{equation}

\section{Calculating $\chiq$ using the PAO basis}\label{formalism}
The perturbative description of the nonlinear optical response to a weak external optical field by means of the second-order optical susceptibility $\chiq$ using the eigenstates and eigenvalues of the tight-binding Hamiltonian $H_0(\bm{k})$ introduced in Section \ref{nonorto} is now described. It is achieved through a perturbative expansion of the density matrix up to second-order in the applied vector potential $\bm{A}(t)$, assumed monochromatic and with no spatial variation. Since the literature on this matter  is abundant~\cite{Cabellos2009,Sipe1993,Aversa1995,Ghahramani1991,Moss1990}, a summary of the main ideas and expressions is provided here; detailed derivations can be found in Appendix~\ref{chi-derivation}.

The second-order susceptibility tensor is decomposed into two contributions \cite{Cabellos2009} $\chiq \equiv A_{ijk}(2\omega)+B_{ijk}(\omega)$, with:
\begin{eqnarray}\label{eq:shg1}
   &\mbox{Im}[A_{ijk}(2\omega)] =\frac{e^3\hbar^3}{2\ve_0 m_e^3}\int_{\text{FBZ}} \frac{d\bm{k}}{(2\pi)^3} \sum_{v=1}^{\textsf{nvb}} \sum_{c=\textsf{nvb}+1}^{\textsf{nbands}} \notag\\
   &\frac{16\pi}{\e_{cv}^{0^3}}\delta(\e^0_{cv}-2\hbar\w)\Bigg(
	\sum_{v'=1}^{\textsf{nvb}}\frac{\mbox{Im}[p_{i;vc} \{p_{j;cv'},p_{k;v'v}\}]}{2\e^0_{cv'}-\e^0_{cv}} -\notag\\
    &\sum_{c'=\textsf{nvb}+1}^{\textsf{nbands}}\frac{\mbox{Im}[p_{i;vc} \{p_{j;cc'},p_{k;c'v}\}]}{2\e^0_{c'v}-\e^0_{cv}} \Bigg),
\end{eqnarray}
and:
\begin{eqnarray}\label{eq:shg2}
    &\mbox{Im}[B_{ijk}(\omega)] =\frac{e^3\hbar^3}{2\ve_0 m_e^3}\int_{\text{FBZ}} \frac{d\bm{k}}{(2\pi)^3} \sum_{v=1}^{\textsf{nvb}}\sum_{c=\textsf{nvb}+1}^{\textsf{nbands}} \notag\\
    &\frac{\pi}{\e_{cv}^{0^3}}\delta(\e^0_{cv}-\hbar\w)\Bigg(
	\sum_{\substack{n=1 \\ (n\neq c)}}^{\textsf{nbands}} \frac{\mbox{Im}[p_{i;nc}\{p_{j;cv},p_{k;vn}\}]}{\e^0_{cn}-2\e^0_{cv}} - \nonumber\\
	&\sum_{\substack{n=1 \\ (n\neq v)}}^{\textsf{nbands}} \frac{\mbox{Im}[p_{i;vn}\{p_{j;nc},p_{k;cv}\}]}{\e^0_{nv}-2\e^0_{cv}} \Bigg),
\end{eqnarray}
where $m_e$ is the electron's mass, $i$, $j$, and $k$ are integers within 1 and 3 denoting Cartesian directions ($x^1\equiv x$, $x^2\equiv y$, or $x^3\equiv z$), $\int d\bm{k}\equiv \int dk_x\int dk_y\int dk_z$, and the integrals run over the  first Brillouin zone with time-reversal symmetry assumed (more on this can be found in section \ref{sec:FBZ_intg}). Sums run over (i) valence states (indices $v$ and $v'$), (ii) conduction states (indices $c$ and $c'$), or (iii) over most of the full range of eigenstates (index $n$ in equation~\eqref{eq:shg2}). (The integer label $k$ above should not be confused with the (real) $k$-point vector $\bm{k}$.) 

Quantities $p_{j;ab}(\bm{k})$ (with $a$ and $b$ representing band indexes) are {\em momentum matrix elements} between Bloch eigenstates (equation~\eqref{kwf2}):
\begin{eqnarray}\label{eq:poperator}
&-\textsf{i}\hbar \Pi_{j;ab}(\bm{k})\equiv p_{j;ab}(\bm{k}) =\int d\bm{r} [\psi_{a\bm{k}}^0(\bm{r})]^*( \hat{p}_j)\psi^0_{b\bm{k}}(\bm{r})\\
&= -\textsf{i}\hbar \int d\bm{r} [\psi_{a\bm{k}}^0(\bm{r})]^* \partial_j \psi_{b\bm{k}}(\bm{r})\nonumber,
\end{eqnarray}
where $\Pi_{j;ab}(\bm{k})$ denotes the matrix element of the (covariant) $\p/\p x^j$ derivative, to be evaluated numerically,
$\partial_j\equiv \frac{\partial}{\partial x^j}$, and $\int d\bm{r}\equiv \int dx\int dy\int dz$.
The procedure for the calculation of these integrals will be given in in section \ref{sec:nonzeroindep}.

The explicit $\bm{k}-$dependence of eigenenergies and linear momentum matrix elements was omitted in equations (\ref{eq:shg1}) and (\ref{eq:shg2}) for the sake of compactness. The curly brackets $\{,\}$ represent a {\it symmetrizing operator} $\{\Xi_j,\Omega_k\}\equiv (\Xi_j\Omega_k+\Xi_k\Omega_j)/2$ that guarantees the intrinsic permutation symmetry $\chi_{ijk}^{\scriptscriptstyle (2)}=\chi_{ikj}^{\scriptscriptstyle (2)}$.

Upon expressing the Bloch functions in terms of equations \eqref{eq:lcao} and \eqref{kwf2}, the matrix element becomes:
\begin{align}\label{puv}
\Pi_{i;ab}(\bm{k}) = \sum_\af\sum_{\af'} &\sum_{n_1}\sum_{n_2}\sum_{n_3} C^{0*}_{\alpha a}(\bm{k})  C^{0}_{\alpha' b}(\bm{k})\times\notag\\
&e^{-\ci\bm{k}\bigdot\bm{R}_{n_1,n_2,n_3}} d_{i;\alpha \alpha'}(\bm{R}_\textsf{p}),
\end{align}
with $\bm{R}_\textsf{p}=\bm{R}_{n_1,n_2,n_3}-\bm{\delta}_{\alpha'} + \bm{\delta}_\alpha$ an implicit function of $n_j$ ($j=1,2,3$), $\alpha$, and $\alpha'$. The integral $d_{i;\alpha \alpha'} (\bm{R}_\textsf{p})$ is defined as:
\begin{equation} \label{d_integrals}
d_{i;\alpha \alpha'} (\bm{R}_\textsf{p}) \equiv\int d\bm{r} \varphi_{\alpha}(\bm{r}-\bm{R}_\textsf{p})\partial_i \varphi_{\alpha'}(\bm{r}),
\end{equation}
and it represents a two-center integral of the derivative operator between the localized pseudo-orbitals $\af$ and $\af^\prime$ separated by $\bm{R}_\textsf{p}$. The sum over $n_1,n_2$ and $n_3$ in equation~\eqref{puv} is restricted such that $|\bm{R}_\textsf{p}|<r_c(\textsf{ce},\textsf{l}(\textsf{ce}),\textsf{ir}(\textsf{l},\textsf{ce}))+r_c(\textsf{ce}',\textsf{l}'(\textsf{ce}'),\textsf{ir}'(\textsf{l}',\textsf{ce}'))$ in figure \ref{radial}, and the granularity of indexes $\af$ and $\af'$ was brought into play again. (Note that as defined in equation (\ref{puv}), the linear momentum in the non-localized basis set is such that a correct definition of the identity follows upon removal of partial derivatives, as equation (\ref{d_integrals}) becomes the overlap matrix $\mathcal{S}$. The equality sign is not written because of the finite size of the basis.)

We rewrite equation~\eqref{puv} for use in section~\ref{results} in the following form. Distances $|\bm{R}_\textsf{p}| = a_{\textsf{p}}$ are ordered increasingly as $0=a_1<a_2<a_3<\cdots a_{\textsf{p}}\cdots$. Then, for each value of $a_{\textsf{p}}$, there exists a set of indices $\alpha$, $\alpha'$, and lattice vectors $\bm{R}_{n_1,n_2,n_3}$ satisfying $|\bm{R}_\textsf{p}| = a_{\textsf{p}}$. The sum in equation~\eqref{puv} is therefore reorganized as:
\begin{align}\label{eq:puv_modf}
\Pi_{i;ab}(\bm{k}) &= \sum_{\textsf{p}=1}^{\textsf{neigh}}\sum_{\substack{\alpha,\alpha' \\ |\bm{R}_\textsf{p}|=a_{\textsf{p}}}} \sum_{\substack{n_1,n_2,n_3\\|\bm{R}_\textsf{p}|=a_{\textsf{p}}}} C^{0*}_{\alpha a}(\bm{k})  C^{0}_{\alpha' b}(\bm{k})\times \notag \\
&e^{-\ci\bm{k}\bigdot\bm{R}_{n_1,n_2,n_3}} d_{i;\alpha \alpha'}(\bm{R}_\textsf{p}),
\end{align}
where the first sum runs over the distinct distances $a_{\textsf{p}}$, and the others run over all indices $\alpha$, $\alpha'$, and $n_1,n_2$ and $n_3$ such that $|\bm{R}_\textsf{p}|=|\bm{R}_{n_1,n_2,n_3}-\bm{\delta}_{\alpha'} + \bm{\delta}_\alpha|=a_{\textsf{p}}$.  The maximum value taken by $\textsf{p}$ is an input variable called \textsf{neigh}. The number of different values of $\bm{R}_{\textsf{p}}$ in the sum above is the {\it number of neighbors} included in the calculation of $Pi_{i;ab}(\bm{k})$ (see section~\ref{results}).

The real part of $\chiq$ is calculated from the imaginary one by means of the Kramers-Kronig relation~\cite{CallawayQTOTSS}:
\begin{align}\label{KK}
\text{Re}[&\chiq]=\frac{1}{\pi}\times \notag\\
&\mathcal{P}\int_{-\infty}^{+\infty}\frac{\text{Im}[\chiqs(-2\w',\w',\w')]}{\w'-\w}d\w',
\end{align}
where $\mathcal{P}$ indicates that the integral is to be evaluated as its principal part~\cite{RileyMMPE}:
\begin{align}\label{Cauchy}
\mathcal{P}\int_{-\infty}^{+\infty}d\w' f(\w') = \lim_{h\rightarrow 0^+}\Big[&\int_{-\infty}^{\w-h}d\w' f(\w')+\notag\\
&\int_{\w+h}^{+\infty}d\w' f(\w')\Big].
\end{align}

\section{Reduction to the irreducible Brillouin zone}\label{sec:FBZ_intg}
The crystal class of a material imposes strict constraints on the allowed nonzero and independent components of its optical susceptibility tensors. By applying each symmetry operation of the point group $G$ (described by matrices $M$) to equations~\eqref{eq:shg1} and~\eqref{eq:shg2}, one arrives at the following homogeneous system of linear algebraic equations:
\begin{equation}\label{nonzeroschi}
	\chiq=M^i_{\;m} M^j_{\;n} M^k_{\;l}\,\chi_{mnl}^{\scp (2)},
\end{equation}
where the tensor-contraction-like structure is used to imply a sum over repeating upper and lower indexes. The solutions of equation~\eqref{nonzeroschi} are documented in the literature~\cite{Boyd,Shen} and will not be reproduced here. Instead, attention is directed to how these symmetries, together with time-reversal symmetry, reduce the integral over the FBZ to only its irreducible part when time-reversal symmetry holds (trIFBZ; see figure~\ref{fig1}(b)). What follows is a slightly modified version of the dyadic approach used in the calculation of the optical response in strained (Si)$_n$/(Ge)$_n$ superlattices~\cite{Ghahramani1991} employing a symmetrizing strategy~\cite{Chadi1973}.

Equations~\eqref{eq:shg1} and~\eqref{eq:shg2} can be written in terms of integrals of invariant quantities with respect to the symmetry operations of the point group $G$ (the energy bands have this property, but the momentum matrix elements do not). These integrals can be  generically written as:
\begin{equation}
    \int_{\text{FBZ}}\frac{d\bm{k}}{(2\pi)^3} f(\bm{k}).
\end{equation}
Since the FBZ has all the symmetries contained in $G$, the equation above can be recast as:
\begin{align}\label{symmed}
    \frac{1}{g}\sum_{i=1}^g\int_{\text{FBZ}} \dk &f(M_i\bm{k})=\notag\\
    &\int_{\text{FBZ}}\dk \frac{1}{g}\sum_{i=1}^g f(M_i\bm{k}).
\end{align}
where $M_i$ is the matrix representation of symmetry operation $i$ in the point group $G$ ($g$ is the order of $G$). The integrand in equation~\eqref{symmed} has the full symmetry of the crystal class, and thus can be reduced to an integral over the irreducible first Brillouin zone (IFBZ) times $g$.

For materials with time-reversal symmetry, a further reduction into the trIFBZ (shown in red in figure \ref{fig1}(b)) can be used, and equation~\eqref{symmed} further reduces to:
\begin{equation}\label{eq:chi_reduced}
    2\int_{\text{trIFBZ}}\dk\sum_{i=1}^g f(M_i\bm{k}).
\end{equation}
Thus:
\begin{equation}\label{eq:p_rot}
	 \Pi_{j;ab}(M_i\bm{k})=M_{\;j;i}^l\Pi_{l;ab}(\bm{k}),
\end{equation}
where indexes $j$ and $l$ represent the entries of the transformation matrix $M_i$. This way, $\chiq$ can be expressed entirely in terms of integrals over the trIFBZ. The final form in equation~\eqref{eq:chi_reduced} depends on the particular crystal class $G$ under consideration, and the results for zinc-blende materials will be presented in section~\ref{results}. In terms of the algorithm, the user inputs \textsf{nk}, leading to a \textsf{nk}$^3$ initial set of points; our algorithm produces a vastly reduced subset of $k-$points to be fed to the SHG solver.

To sample the trIFBZ explicitly, we set up a regular grid in $k$-space with size $8 \text{ \AA}^{-3}/\textsf{nk}^3$: $\bm{k}(j_x,j_y,j_z)=2\text{ \AA}^{-1}(j_x-1,j_y-1,j_z-1)/\textsf{nk}$, with $j_x$, $j_y$, $j_z$ and \textsf{nk} integers. Imposing the condition that $\bm{k}{j_x j_y j_z}$ lies inside the trIFBZ, we determine the number of actual $k$-points to be evaluated. Using the reciprocal vectors $\bm{b}_j$, the corresponding numbers of actual $k-$points used for \textsf{nk} = 20, 30, and 35 are 356, 1078, and 1589, respectively.

\section{Symmetries of two-center integrals among PAOs}\label{two_center_intg}

The symmetries of the crystal determine the optimized trIFBZ. We now discuss the symmetries of the two-center matrix elements of the linear momentum. Integrals of this kind have been previously calculated by reducing them to integrals of the position operator in between the PAOs $\varphi_\alpha(\bm{r})$~\cite{Moss1986}, although the most common approach is to perform a Fourier transformation and to take advantage of the convolution-like  form of equation~\eqref{d_integrals}~\cite{Sankey1989}.

\begin{figure}[tb]
	\begin{center}
		\includegraphics[width=\columnwidth]{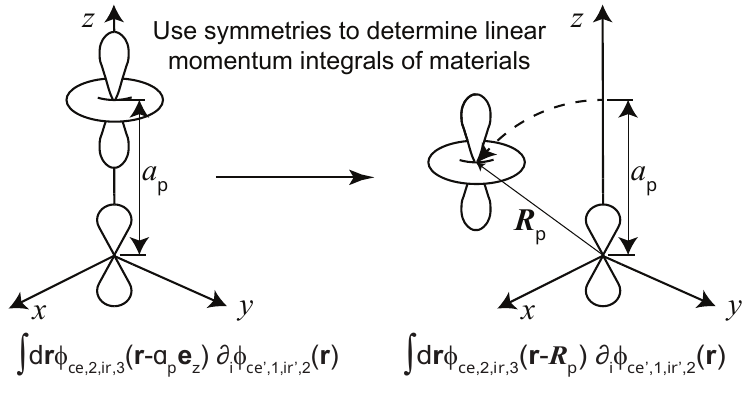}
		\caption{The heart of this development is to calculate linear momentum integrals among PAOs in highly-symmetric configurations, and to use symmetries to determine actual integrals on materials.\label{scheme-rot}}
	\end{center}
\end{figure}

In this work, the integration of equation~\eqref{d_integrals} is done by rotating the Cartesian axes in such a way that the new $z'$ axis is brought into the $\bm{R}_\textsf{p}$ direction, something akin to the Slater-Koster method~\cite{Slater1954,Sankey1989}. In this approach (schematically depicted in figure \ref{scheme-rot}), the spatial symmetries of the real spherical harmonics imply that many of the integrals $d_{i;\alpha\alpha'}(a_{\textsf{p}}\hat{\bm{z}})$, where $a_{\textsf{p}}$ is the magnitude of $\bm{R}_\textsf{p}$, either vanish or are related by symmetry. Consequently, only a small subset of them must be evaluated explicitly. This conceptualization sets our methodology apart from that described in Ref.~\cite{Li2022}, which is reliant on a localized basis as well. The procedure is as follows.

If the vector $\bm{R}_\textsf{p}$ is written in spherical coordinates as $\bm{R}_\textsf{p}=a_{\textsf{p}}(\cos\phi\sin\theta,\sin\phi\sin\theta,\cos\theta)$, the angles $\phi$ and $\theta$ are the Euler angles (the angle $\gamma$ can be arbitrarily set to zero). The orthogonal matrix executing this axial rotation is:
\begin{equation}\label{rotation}
	P(\phi,\ti)=\left(\begin{array}{ccc}
		\cos\phi\cos\ti & \sin\phi\cos\ti & -\sin\ti \\
		-\sin\phi &  \cos\phi & 0 \\
	    \cos\phi\sin\theta         &  \sin\phi\sin\theta        & \cos\theta
	\end{array} \right).
\end{equation}
With this, we have $P \bm{R}_\textsf{p}=\tilde{\bm{R}}_p$, where $\tilde{\bm{R}}_p=(0,0,a_{\textsf{p}})$. In components, this can be written as:
\begin{equation}
P \bm{R}=\tilde{\bm{R}}\to P_{\;i}^{j} R^i=\tilde{R}^j,
\end{equation}
where $\tilde{R}^j$ are the components in the rotated frame, and a sum over repeated indexes appearing at both covariant and a contravariant locations continues to be implied. The transformation of the gradient follows from this being a covariant object:
\begin{equation}\label{eq:chain_rule}
P^{-1} \boldsymbol{\nabla}=
\tilde{\boldsymbol{\nabla}}\to (P_i^j)^{-1}\partial_i=
(P_i^j)^{T}\partial_i=
P_j^i\partial_i=
\tilde{\partial_j}.
\end{equation}

The arguments $\phi$ and $\theta$ appearing in the rotation matrix $P(\phi,\theta)$ will be omitted in what follows. Since the integration in equation~\eqref{d_integrals} is over all space, the substitution $\bm{r}\rightarrow P^T\bm{r}$ effects a rotation:
\begin{equation}\label{eq:Pij_rotated}
d_{i;\af\af'}(\bm{R}_\textsf{p})=P^l_{\;i}\dv\vphi_\af(P^T(\bm{r}-\bm{\tilde{R}}_p))\p_l\vphi_{\af'}(P^T\bm{r}),
\end{equation}
where the partial derivative has been transformed according to the chain rule (equation~\eqref{eq:chain_rule}) and, as before, the subscripts $\af$ and $\af'$ are understood to represent $(\fce,\li,\fir,\fim)$ and $(\fce',\li',\fir',\fim')$, respectively.

General expressions of the kind $\varphi_\mu(P\bm{r})$, where $\varphi_\mu$ is a linear combination of PAOs, can be worked out from the $D^{(\textsf{l})}(\phi,\ti,0)$ rotation matrices for spherical harmonics in their complex form~\cite{TinkhamGTaQM}. The corresponding matrices for spherical harmonics in their real form (equation~\eqref{eq:RealHarmonics})  can be readily obtained from $D^{(\textsf{l})}(\phi,\ti,0)$, and both kind of matrices can be found in Appendix~\ref{Dmatrices} for $p-$ and $d-$orbitals ($s-$orbitals do not require special treatment due to their spherical symmetry). Since the radial part of PAO is invariant under rotations, these matrices apply equally to the PAOs in Table~\ref{ta:PAOs} as to spherical harmonics.

The explicit forms of the rotated PAOs in equation~\eqref{eq:Pij_rotated} are better analyzed by using the full notation $\vphi_{\fce,\li,\fir,\fim}(\bm{r})$ for the PAOs introduced in section \ref{nonorto}. Using the the matrix in equation~\eqref{rotationp},  the rotation properties for PAOs with $p-$character ($\li=1$) are as follows ($j,l=1,2,3$):
\begin{equation}
    \varphi_{\textsf{ce},1,\textsf{ir},j}(P^T\bm{r})=
M^l_{\;j;1}\varphi_{\textsf{ce},1,\textsf{ir},l}(\bm{r}),
\end{equation}
with $M_1$ listed as equation \eqref{rotationp} being different to the spatial rotation matrix $P$ (equation \eqref{rotation}) from the conventional orbital designation in Table \ref{ta:SphericalHarmonics} \cite{VarshalovichQTAM} (rotation angles were omitted from the matrix $M_1$ for compactness).

Due to convention, PAOs containing $d-$channels (\textsf{l}=2) were sequentially listed as $\{d_{xy},d_{yz},d_{3z^2-r^2},d_{zx},d_{x^2-y^2}\}\to \{1,2,3,4,5\}$ in Table \ref{ta:SphericalHarmonics}. PAOs containing these angular channels transform as follows ($j,l=1,\cdots,5$):
\begin{equation}
    \varphi_{\textsf{ce},2,\textsf{ir},j}(P^T\bm{r})=M^l_{\;j;2}\,\varphi_{\textsf{ce},2,\textsf{ir},l}(\bm{r}),
\end{equation}
the angular arguments in $M_2$ will be omitted from now on, as it was done with $M_1$.  Using equations~\eqref{rotation},~\eqref{eq:Pij_rotated},~\eqref{rotationp} and~\eqref{rotationd}, the reduction of equation~\eqref{d_integrals} into combinations of similar integrals along the $z-$axis is better understood by considering all possibilities.

(i) Integrals between PAOs with $s-$angular momentum ($\li=0$):
\begin{eqnarray}\label{ss}
&\dv\varphi_{\textsf{ce},0,\textsf{ir},0}(\bm{r}-\bm{R}_\textsf{p})\p_i \varphi_{\textsf{ce}',0,\textsf{ir}',0}(\bm{r})=\nonumber\\
&P^l_{\;i}\int d\bm{r} \varphi_{\textsf{ce},0,\textsf{ir},0}(\bm{r}-a_{\textsf{p}}\bm{e}_z)\p_l \varphi_{\textsf{ce}',0,\textsf{ir}',0}(\bm{r}).
\end{eqnarray}
The unprimed PAO lies a distance $a_{\textsf{p}}$ above the primed PAO on the expression below the equality sign in equation~\eqref{ss}.

(ii) Integrals between PAOs of $s-$ and $p-$character, respectively ($j,k=1,2,3$):
\begin{eqnarray}\label{spd}
&\int d\bm{r} \varphi_{\textsf{ce},0,\textsf{ir},1}(\bm{r}-\bm{R}_\textsf{p})\partial_i \varphi_{\textsf{ce}',1,\textsf{ir}',j}(\bm{r})=\nonumber\\
&P_{\;i}^lM_{\;j;1}^k\int d\bm{r} \varphi_{\textsf{ce},0,\textsf{ir},1}(\bm{r}-a_{\textsf{p}}\bm{e}_z)\partial_l \varphi_{\textsf{ce}',1,\textsf{ir}',k}(\bm{r}).
\end{eqnarray}
A similar equation holds when the PAO centered at the origin has $d-$type orbital angular momentum ($j,k=1,\cdots,5$):
\begin{align}
&\int d\bm{r} \varphi_{\textsf{ce},0,\textsf{ir},0}(\bm{r}-\bm{R}_\textsf{p})\partial_i \varphi_{\textsf{ce}',2,\textsf{ir}',j}(\bm{r})=\notag\\
&P_{\;i}^lM_{\;j;2}^k\int d\bm{r} \varphi_{\textsf{ce},0,\textsf{ir},0}(\bm{r}-a_{\textsf{p}}\bm{e}_z)\partial_l \varphi_{\textsf{ce}',2,\textsf{ir}',k}(\bm{r}).
\end{align}
(iii) Integrals between PAOs with angular momentum \textsf{l}=1 ($j,k,l,n=1,2,3$):
\begin{align}\label{pp}
&\int d\bm{r} \varphi_{\textsf{ce},1,\textsf{ir},j}(\bm{r}-\bm{R}_\textsf{p})\partial_i \varphi_{\textsf{ce}',1,\textsf{ir}',k}(\bm{r})=\\
&M_{\;j;1}^l P_{\;i}^m M_{\;k;1}^n\dv\varphi_{\textsf{ce},1,\textsf{ir},l}(\bm{r}-a_{\textsf{p}}\bm{e}_z)\partial_m \varphi_{\textsf{ce}',1,\textsf{ir}',n}(\bm{r}).\notag
\end{align}
A similar expression holds true for integrals between PAOs with $d-$character  using $M_2$ instead ($j,k,l,n=1,\cdots,5$):
\begin{align}\label{pp}
&\dv\varphi_{\textsf{ce},2,\textsf{ir},j}(\bm{r}-\bm{R}_\textsf{p})\partial_i \varphi_{\textsf{ce}',2,\textsf{ir}',k}(\bm{r})= \\
&M_{\;j;2}^l P_{\;i}^m M_{\;k;2}^n\dv\varphi_{\textsf{ce},2,\textsf{ir},l}(\bm{r}-a_{\textsf{p}}\bm{e}_z)\partial_m \varphi_{\textsf{ce}',2,\textsf{ir}',n}(\bm{r}).\notag
\end{align}
(iv) Integrals between $p-$ ($\li=1$) and $d-$orbitals ($\li=2$) have the following form ($j=1,2,3$, $k=1,\cdots,5$):
\begin{align}\label{pd}
&\int d\bm{r} \varphi_{\fce,1,\fir,j}(\bm{r}-\bm{R}_\textsf{p})\p_i \varphi_{\fce',2,\fir',k}(\bm{r})= \\
&M_{\;j;1}^l P_{\;i}^m M_{\;k;2}^n\int d\bm{r} \varphi_{\textsf{ce},2,\textsf{ir},l}(\bm{r}-a_{\textsf{p}}\bm{e}_z)\partial_m \varphi_{\textsf{ce}',2,\textsf{ir}',n}(\bm{r}).\notag
\end{align}
Cases where the roles of the orbitals are swapped can be calculated by using the identity ($\af\equiv(\fce,\li,\fir,\fim)$ and $\af'\equiv(\fce',\li',\fir',\fim')$):
\begin{equation}\label{par}
	d_{j;\af\af'}(\bm{R}_\textsf{p})=\mathcal{P}_\alpha \mathcal{P}_{\alpha'} \,d_{j;\af'\af}(\bm{R}_\textsf{p}),
\end{equation}
which comes from the finite range of the PAOs. Here  $\mathcal{P}_\alpha$ is the parity of the $\varphi_\alpha$ PAO (1 for $s-$ and $d-$orbitals, and $-1$ for $p-$orbitals). This identity has been verified in our numerical implementation.

\subsection{Nonzero independent integrals}\label{sec:nonzeroindep}
After bringing the general two-center integrals $-$equation~\eqref{d_integrals}$-$into linear combinations of terms like $\int d\bm{r}\varphi_\alpha(\bm{r}-a_{\textsf{p}}\bm{e}_z)\partial_l\varphi_{\alpha'}(\bm{r})$, it must be noted that many of these integrals are identically zero, and the nonzero ones are not fully independent. These results--which reduce the number of numerical integrations--follow from symmetry considerations and are summarized in what follows. To continue reducing notation, we remove the dependence in position within the integrals, understanding that the PAO before the partial derivative is located at $a_{\textsf{p}}\bm{e}_z$, and the one after the derivative is centered at the origin. Also, when writing partial derivatives, the notation $\p_1=\p_x$, $\p_2=\p_y$ and $\p_3=\p_z$ will be used.

For the case in equation~\eqref{ss}, where both $\vphi_\af$ and $\vphi_{\af'}$ have $s-$dependency, we have:
\begin{align}\label{nonzer-ss}
	&\dv\vphi_{\fce,0,\fir,1}\p_3 \vphi_{\fce',0,\fir',1} \neq 0, \text{and} \\
    &\dv\vphi_{\fce,0,\fir,1}\p_1 \vphi_{\fce',0,\fir',1} = \dv\vphi_{\fce,0,\fir,1}\p_2 \vphi_{\fce',0,\fir',1} \neq 0, \notag
\end{align}
thus, only one nonzero element exists for each pair of $s-$orbitals in the calculation, for $N_s(N_s+1)/2$ integrals in total, if $N_s$ is the total number of $s-$orbitals involved (counting all the different $\fce$, $\fce'$, $\fir$ and $\fir'$ possibilities). Equation~\eqref{par} takes care of the remaining $N_s(N_s-1)/2$ integrals between different $s$ PAOs.

In the case of integrals between $p-$ ($\li=1$) and $s-$orbitals ($\li=0$), the expressions take the form:
\begin{align}\label{nonzer-ss}
	&\dv\vphi_{\fce,1,\fir,2}\p_3 \vphi_{\fce',0,\fir',1} \neq 0, \text{and} \\
    &\dv\vphi_{\fce,1,\fir,3}\p_1 \vphi_{\fce',0,\fir',1} = \dv\vphi_{\fce,1,\fir,1}\p_2 \vphi_{\fce',0,\fir',1} \neq 0, \notag
\end{align}
all the other integrals being zero. As before, the total number of integrals of this type are obtained by considering the total number of PAOs with $s-$ and $p-$character.  Between $d-$  and $s-$orbitals, the nonzero independent contributors are:
\begin{align}\label{nonzer-ss}
	&\dv\vphi_{\fce,2,\fir,3}\p_3 \vphi_{\fce',0,\fir',1} \neq 0,  \\
    &\dv\vphi_{\fce,2,\fir,5}\p_3 \vphi_{\fce',0,\fir',1} \neq 0,  \notag\\
    &\dv\vphi_{\fce,2,\fir,4}\p_1 \vphi_{\fce',0,\fir',1} = \dv\vphi_{\fce,2,\fir,2}\p_2 \vphi_{\fce',0,\fir',1} \neq 0. \notag
\end{align}
There are only three independent integrals between PAOs with $p-$orbital character:
\begin{align}
&\dv\vphi_{\fce,1,\fir,2}\p_3 \vphi_{\fce',1,\fir',2} \neq 0,\notag\\
&\dv\vphi_{\fce,1,\fir,2}\p_1 \vphi_{\fce',1,\fir',3} = \dv\vphi_{\fce,1,\fir,2}\p_2 \vphi_{\fce',1,\fir',1}\neq 0,\label{pp1}\\
&\dv\vphi_{\fce,1,\fir,3}\p_1 \vphi_{\fce',1,\fir',2} =
\dv\vphi_{\fce,1,\fir,1}\p_2 \vphi_{\fce',1,\fir',2}= \notag\\
&-\dv\vphi_{\fce,1,\fir,1}\p_3 \vphi_{\fce',1,\fir',1} = -\dv\vphi_{\fce,1,\fir,3}\p_3 \vphi_{\fce',1,\fir',3}.\label{pp2}
\end{align}
If $\fce=\fce'$ and $\fir=\fir'$, a further symmetry appears, which makes integrals in equations~\eqref{pp1} and~\eqref{pp2} the same, thus reducing the number of independent integrals to just two. For integrals of $p-$ and $d-$orbitals there are only eleven nonzero ones, but only five of them are independent:
\begin{align}
&\dv\vphi_{\fce,1,\fir,2}\p_3 \vphi_{\fce',2,\fir',3} \neq 0,\notag\\
&\dv\vphi_{\fce,1,\fir,1}\p_1 \vphi_{\fce',2,\fir',1} =
 \dv\vphi_{\fce,1,\fir,3}\p_2 \vphi_{\fce',2,\fir',1} \neq 0,\notag\\
&\dv\vphi_{\fce,1,\fir,3}\p_1 \vphi_{\fce',2,\fir',3} =
 \dv\vphi_{\fce,1,\fir,1}\p_2 \vphi_{\fce',2,\fir',3} \neq 0, \notag\\
&\dv\vphi_{\fce,1,\fir,3}\p_1 \vphi_{\fce',2,\fir',5} =
-\dv\vphi_{\fce,1,\fir,1}\p_2 \vphi_{\fce',2,\fir',5} \neq 0,\notag\\
&\dv\vphi_{\fce,1,\fir,2}\p_1 \vphi_{\fce',2,\fir',4} =
 \dv\vphi_{\fce,1,\fir,1}\p_2 \vphi_{\fce',2,\fir',2} = \notag\\
&-\dv\vphi_{\fce,1,\fir,1}\p_3 \vphi_{\fce',2,\fir',2} =
-\dv\vphi_{\fce,1,\fir,3}\p_3 \vphi_{\fce',2,\fir',4}.
\end{align}
The $d-d$ integrals have a more complicated structure. In order to simplify the presentation, the notation $\mathcal{J}^{\fce,\fir}_{\fce',\fir'}(\fim,\fim',i)$ is introduced:
\begin{equation}
    \mathcal{J}^{\fce,\fir}_{\fce',\fir'}(\fim,\fim',i)\equiv \dv\vphi_{\fce,2,\fir,\fim}(\bm{r}-a_{\textsf{p}}\bm{e}_z)\partial_i \vphi_{\fce',2,\fir',\fim'}(\bm{r}).
\end{equation}

The nonzero two-center integrals in this case are:
\begin{align}
&\mathcal{J}^{\fce,\fir}_{\fce',\fir'}(3,3,3)\neq0, \notag\\
&\mathcal{J}^{\fce,\fir}_{\fce',\fir'}(2,2,3)=\mathcal{J}^{\fce,\fir}_{\fce',\fir'}(4,4,3)\neq0, \notag\\
&\mathcal{J}^{\fce,\fir}_{\fce',\fir'}(3,4,1)=\mathcal{J}^{\fce,\fir}_{\fce',\fir'}(2,3,2)\neq0, \notag\\
&\mathcal{J}^{\fce,\fir}_{\fce',\fir'}(1,2,1)=\mathcal{J}^{\fce,\fir}_{\fce',\fir'}(5,4,1)=\mathcal{J}^{\fce,\fir}_{\fce',\fir'}(1,4,2)=\notag\\
&\mathcal{J}^{\fce,\fir}_{\fce',\fir'}(2,1,1)=\mathcal{J}^{\fce,\fir}_{\fce',\fir'}(4,5,1)= \mathcal{J}^{\fce,\fir}_{\fce',\fir'}(4,1,2)=\notag\\
&-\mathcal{J}^{\fce,\fir}_{\fce',\fir'}(5,2,2)=-\mathcal{J}^{\fce,\fir}_{\fce',\fir'}(1,1,3)=-\mathcal{J}^{\fce,\fir}_{\fce',\fir'}(5,5,3)=\notag\\
&-\mathcal{J}^{\fce,\fir}_{\fce',\fir'}(2,5,2)\neq0.
\end{align}

\begin{figure}[tb]
	\centering
	\includegraphics[width=0.9\columnwidth]{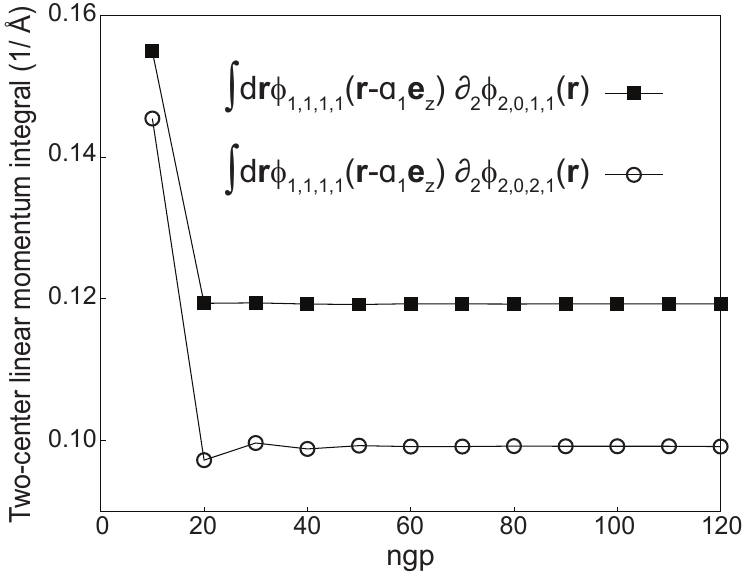}
	\caption{Examples of two-center integrals (equation~\eqref{d_integrals}), as a function of the number of integration points \textsf{ngp}$^3$, which is another input in our code ($a_1=1.911$ \AA). Convergence of the order of $10^{-3}$ is achieved for $\textsf{ngp}=60$.\label{two-center}}
\end{figure}

For the few integrals that have to be numerically evaluated, the Gauss-Legendre quadrature with a fine grid containing \textsf{ngp}$^3$ points over a rectangular parallelepiped centered at the point $(0,0,a_{\textsf{p}})$ with ranges $-a_M\moi x\moi a_M$, $-a_M\moi y\moi a_M$ and $-a_M\moi z\moi a_{\textsf{p}}+ a_M$ was used, where $a_M$ is a length above the largest cutoff radius $a_M > \text{max}(r_c(\textsf{ce},\textsf{l}(\textsf{ce}),\textsf{ir}(\textsf{l},\textsf{ce})))$; {\em i.e.}, larger than the maximum spatial extension of the PAOs, to guarantee an appropriate convergence. Figure~\ref{two-center} depicts integrals involving the $\vphi_{1,1,1,1,}(\bm{r}-a_{\textsf{p}}\bm{e}_z)$ PAO of silicon and the $\vphi_{2,0,\fir,1}(\bm{r})$ ($\fir=1,2$) PAOs of carbon as a function of \textsf{ngp}. Convergence is achieved when \textsf{ngp}$\geqslant 60$.

\begin{figure}[tb]
    \centering
    \includegraphics[width=0.95\linewidth]{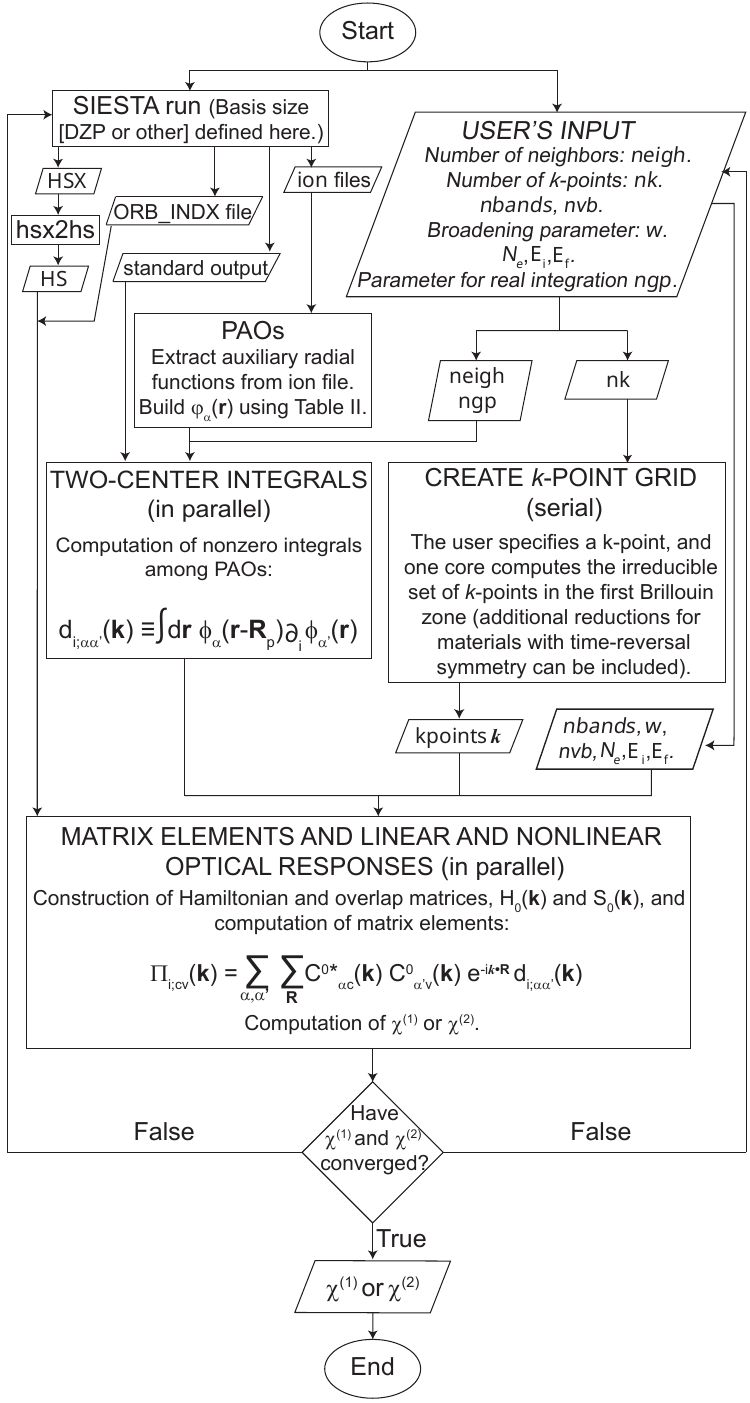}
    \caption{Flowchart showing the step-by-step procedure used in this paper to obtain the second-order optical susceptibility.}
    \label{fig:flowchart}
\end{figure}

\section{Computational details}\label{sec:comp_details}
The Fortran code ran on a single Dell R640  compute node consisting of two Intel Xeon Gold 6130 processors, with 32 cores in total, and 192 GB of memory.  We used the Intel Fortran compiler version 2021.7.0 and Intel MKL numerical linear algebra libraries (version 22.1.0) for the eigenvalue solver. We used Intel MPI 21.8.0 libraries as well.

Previous sections have been written with an algorithmic mindset; these already provide multiple details of the tool we developed. Figure \ref{fig:flowchart} unveils the workflow adopted in this work. The process begins with a SIESTA self-consistent calculation with a target PAO basis size and with home-made pseudopotentials \cite{Rivero2015}, from which output files alluded to earlier in the paper are extracted: files with \verb|ion| extension, \verb|HS|, a file with \verb|ORB_INDX| extension, and the standard \verb|output|. Table \ref{ta:PAOs} is used to construct the PAOs (section \ref{nonorto}). With these numerical PAOs the nonzero two-center integrals enumerated in section~\ref{sec:nonzeroindep} are calculated.

The user specifies \textsf{nk} for the initial $k$-point sampling. A single core generates the corresponding set of \textit{k}-points $\bm{k}$ in the trIFBZ and writes them to a file named \verb|kpoints| (section \ref{sec:FBZ_intg}).

The user also selects the number of nearest neighbors distances to include \textsf{neigh} (see equation~\eqref{eq:puv_modf}), as well as the total bands \textsf{nbands} to compute. The nonzero two-center integrals defined in equation~\eqref{d_integrals} (section \ref{two_center_intg}) are then evaluated in parallel.

Using the \verb|HS| file, the generalized eigenvalue problem is then solved on the \textit{k}-grid specified in \verb|kpoints| to obtain the unperturbed eigenvalues and eigenvectors (section \ref{nonorto}). This step is parallelized as well. After setting a value for the broadening \textsf{w}, the energy step (number of energy points) $\textsf{N}_\textsf{e}$, and the energy bounds $\textsf{E}_\textsf{i}$ and $\textsf{E}_\textsf{f}$, the linear and nonlinear optical susceptibilities (section \ref{formalism}) are subsequently assembled. Convergence with respect to input parameters is verified afterwards. The first- and second-order susceptibilities are written into text files.

\section{Results}\label{results}
Cubic (3C) SiC is one of the many allotropes~\cite{Harris1995} of silicon carbide and has a zinc-blende structure with a fcc crystal system spanned by lattice vectors $\mathbf{a}_1=a(\bm{e}_x+\bm{e}_y)/2$, $\mathbf{a}_2=a(\bm{e}_y+\bm{e}_z)/2$ and $\mathbf{a}_3=a(\bm{e}_z+\bm{e}_x)/2$, $a$ is the side of the cubic conventional unit cell and the lattice parameter is $a/\sqrt{2}$ (figure~\ref{fig1}(a)). The primitive cell contains one Si atom at the origin and a C atom at $(\mathbf{a}_1+\mathbf{a}_2+\mathbf{a}_3)/4$. This polymorph belongs to the F43m space group (space group No. 216) with a point group $\bar{4}3m$ or $T_d$ (group of symmetry transformations of a regular tetrahedron~\cite{LandauQMNRT}) that have $g=24$ symmetry operations, as listed in Appendix~\ref{sec:reduc_BZ}. The reciprocal space is a body centered cubic lattice, spanned by lattice vectors $\mathbf{b}_1$, $\mathbf{b}_2$ and $\mathbf{b}_3$. The FBZ is a truncated octahedron, and it is shown in figure~\ref{fig1}(b), along with high symmetry points $\Gamma$, $K$, $W$, $X$, $U$ and $L$~\cite{Setyawan2010}. The trIFBZ (1/48 the volume of the FBZ) is also shown.

Density functional theory (DFT) calculations were carried out using the siesta-trunk-462 version of SIESTA, employing the generalized gradient approximation (GGA) with the Perdew–Burke–Ernzerhof (PBE)~\cite{Perdew1996} exchange–correlation functional. A Monkhorst–Pack~\cite{Monkhorst1976} grid of $12 \times 12 \times 12$ $k$-points centered at $\Gamma$ was used, together with default values of $100$~Ry for the \verb|MeshCutOff| and $0.01$~Ry for the \verb|PAO.EnergyShift| input parameters. The optimal lattice constant $a$ is $4.414\,$\AA.

Calculations were performed using single-$\zeta$ plus polarization (\verb|SZP|), double-$\zeta$ plus polarization (\verb|DZP|), and triple-$\zeta$ plus polarization (\verb|TZP|) basis sets (calculations with \verb|SZP| and \verb|DZP| basis sets proceeded on a structure optimized with the \verb|DZP| basis set). The indirect band gap ($\Gamma\!-\!X$) is in fair agreement with the $1.29$~eV gap reported in  Ref.~\cite{Wu2008}. The values obtained here are $1.45$, $1.42$, and $1.40$~eV, for \verb|SZP|, \verb|DZP| and \verb|TZP| basis, respectively. No scissor operator was used to adjust the band gap~\cite{Cabellos2009}.
\begin{figure}[tb]
	\centering	\includegraphics[width=\columnwidth]{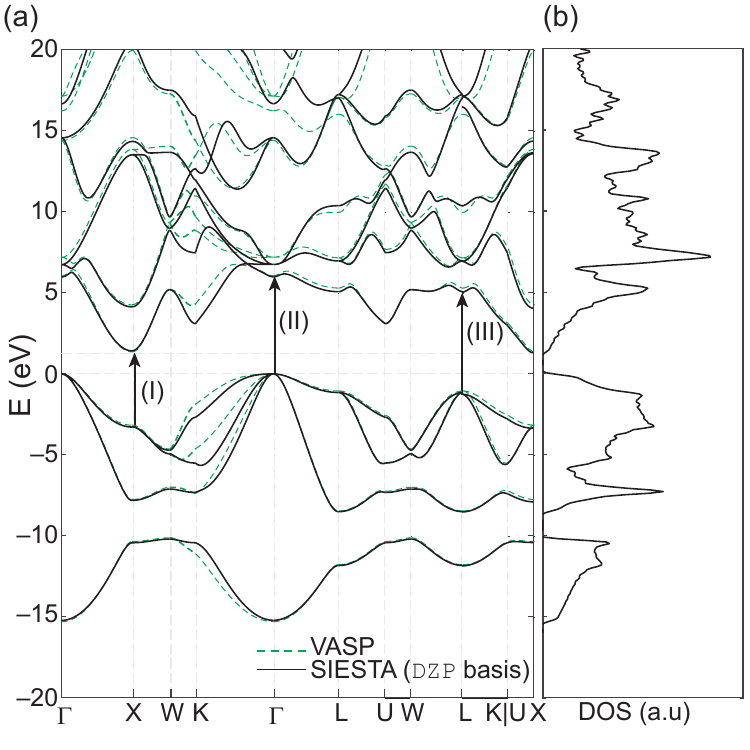}
	\caption{(a) Energy bands of 3C-SiC along high symmetry paths shown in figure~\ref{fig1}(b) for a VASP calculation (green dashed lines) and a SIESTA calculation with a double-$\zeta$ plus polarization (\texttt{DZP}, black solid line) basis. Three direct transitions are highlighted. (b) Density of states (DOS) in arbitrary units.\label{fig:bands}}
\end{figure}

Figure~\ref{fig:bands}(a) shows the band structure calculated with the SIESTA DFT package using the DZP basis set (black solid line), alongside results obtained with the (plane-wave) VASP code (green dashed line) \cite{Hafner2008}. The density of states (DOS) shown in figure~\ref{fig:bands}(b) was calculated using the linear tetrahedral method~\cite{Lehmann1972}.

\begin{figure*}[tb]
    \centering
    \includegraphics[width=\textwidth]{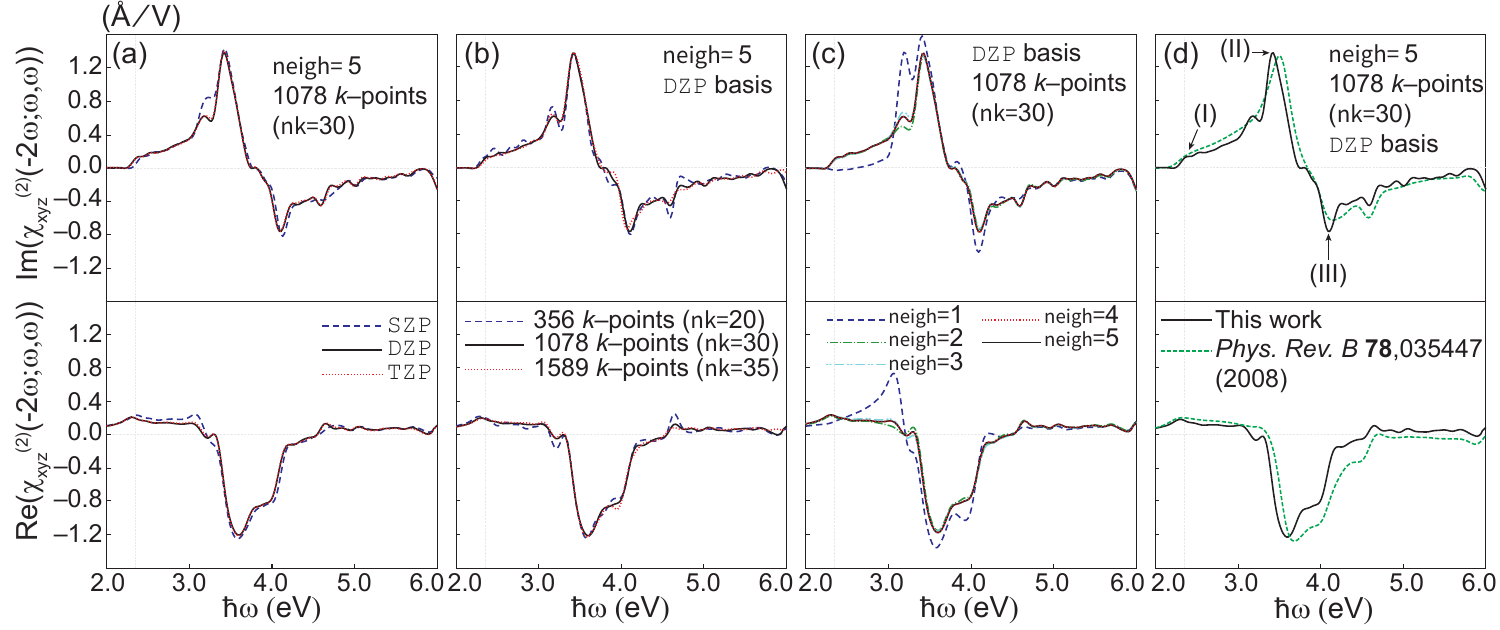}
    \caption{$\text{Im}[\xyz]$ as a function of the photon energy $\hw$ for (a) three different samplings of the trIFBZ and a fixed number of neighbors and basis set (\texttt{DZP}); (b) three different basis sets, with a fixed number of neighbors and samplig of the trIFBZ; and (c) five different number of neighbors and fixed basis set (\texttt{DZP}) and sampling of the trIFBZ. (d) Comparison with previous literature. Vertical dashed lines indicating the onset of the susceptibility are shown in all panels.}
    \label{fig:convergence}
\end{figure*}

The imaginary parts of the first- and second-order optical susceptibilities, given by equations~\eqref{eq:chi_ij}, \eqref{eq:shg1}, and \eqref{eq:shg2}, were evaluated using a Gaussian broadening scheme. The Dirac delta is written as $\delta(\e_{cv}-s\hw)=\delta(\textsf{w}\times(\e_{cv}-s\hw)/\textsf{w})=(1/\textsf{w})\delta((\e_{cv}-s\hw)/\textsf{w})$, where $s=1,2$, and \textsf{w} is a positive energy making the argument in the Dirac delta dimensionless and giving its broadening. The following approximation is used~\cite{Methfessel1989}:
\begin{equation}\label{eq:Dirac}
\delta(x) \simeq \sum_{n=0}^N A_n H_{2n}(x) e^{-x^2},
\end{equation}
where $H_{2n}$ are the Hermite polynomials of even order and $A_n=(-1)^n/(n!, 4^n\sqrt{\pi})$. We have verified that setting $\textsf{w}=0.2\,$eV and $N=2$, convergence in both $A_{ijk}(2\w)$ and $B_{ijk}(\w)$ is achieved. This is the order of magnitude of the broadening most commonly used in the literature~\cite{Guo2004}.

In our calculations, we have $\chiq$ for real frequencies only, so we transform equation~\eqref{KK} into:
\begin{align}\label{eq:KK_final}
\text{Re}[&\chiq]=\frac{2}{\pi}\times\notag\\
&\mathcal{P}\int_0^{+\infty}d\w'\frac{ \w'}{\w'^2-\w^2}\text{Im}[\chiqs(-2\w',\w',\w')],
\end{align}
and use $h=(\textsf{E}_\textsf{f}-\textsf{E}_\textsf{i})/\textsf{N}_\textsf{e}$ in equation~\eqref{Cauchy} to be the energy bin ($\textsf{E}_\textsf{i}$ and $\textsf{E}_\textsf{f}$ are the energy bounds over which the susceptibilities are calculated, and are also defined by the user). Other choices do not change the real part significantly. The integration in equation~\eqref{eq:KK_final} is performed by a trapezoidal rule using the same energy mesh used in calculating $\text{Im}[\chiq]$ without including the singularity at $\hw$.


\subsection{Second-order susceptibility}

The only nonzero components of $\chiq$, allowed by the point group $T_d$ of 3C-SiC, are all equal and can be worked out from equation~\eqref{nonzeroschi}:
\begin{equation}\label{eq:nonzero_chiq}
	\chi_{xyz}^{\scriptscriptstyle (2)}=\chi_{yzx}^{\scriptscriptstyle (2)}=\chi_{zxy}^{\scriptscriptstyle (2)},
\end{equation}
and $\chi^{\scriptscriptstyle (2)}_{ijk}=\chi^{\scriptscriptstyle (2)}_{ikj}$. By applying the 24 operations of the point group $T_d$ to the sum over $M$ in equation~\eqref{eq:chi_reduced}, we get after a lengthy calculation the following final form for $\xyz$ (see Appendix~\ref{sec:reduc_BZ} for details):
\begin{equation}
    \xyz(\w)=16\big( \tilde{\chi}^{\scp (2)}_{xyz}+\tilde{\chi}^{\scp (2)}_{yzx}+\tilde{\chi}^{\scp (2)}_{zxy} \big).
\end{equation}
The notation $\tilde{\chi}^{\scp (2)}_{ijk}$ is to be understood as the sum of the two integrals in equations~\eqref{eq:shg1} and~\eqref{eq:shg2} but {\it over the} trIFBZ {\it only}, instead if the whole FBZ, as it is apparent from equation~\eqref{eq:chi_reduced}. This is the expression we use in our numerical implementation. 

The numerical evaluation of $\chiq$ requires a thorough set of convergence tests. Beyond the conventional test involving the density of the \textit{k}-point grid, our approach also demands convergence with respect to \textsf{neigh} and  to the PAO basis set size. The ability to independently tune these parameters is one of the main advantages of this methodology, as it allows us to systematically identify an optimal approximation.

\begin{figure}[tb]
    \centering
    \includegraphics[width=\linewidth]{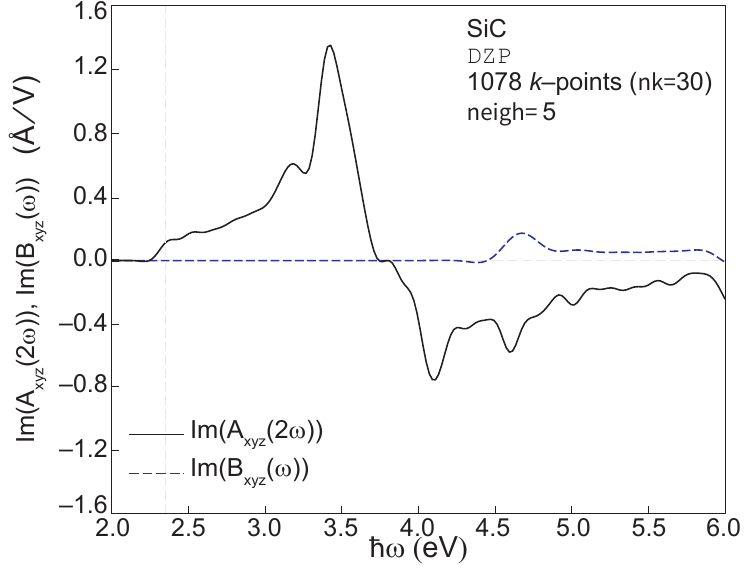}
    \caption{Contributions $\text{Im}[A_{xyz}(2\w)]$ and $\text{Im}[B_{xyz}(\w)]$ to $\text{Im}[\xyz]$ of 3C-SiC as a function of the photon energy $\hw$. \texttt{DZP} calculation with with $1078$ $k-$points, $\textsf{w}=0.2\,$eV, and $N=2$. }
    \label{fig:ABxyz}
\end{figure}
The results of these convergence tests are presented in figure~\ref{fig:convergence}.  All tests were performed using $\verb|nbands|=18$, a Gaussian broadening of $\verb|w|=0.2$ eV, and an energy mesh of $\verb|Ne|=200$ points. Figure~\ref{fig:convergence}(a) displays the real and imaginary components of $\xyz(-2\omega;\omega,\omega)$ computed using a fixed set of 46 neighbors and 1078 \textit{k}-points, while varying the basis from \verb|SZP| to \verb|DZP| and \verb|TZP|. As expected, results based on a  \verb|SZP| PAO basis set are the least accurate, whereas those obtained with the \verb|DZP| basis can be hardly told appart from those using a lager, \verb|TZP| basis: we work with \verb|DZP| sets from now on.

Figure~\ref{fig:convergence}(b) examines convergence with respect to the sampling of the trIFBZ using a fixed \verb|DZP| basis and 46 neighbors (\textsf{neigh}=5). The comparison among three \textit{k}-point grids demonstrates that 1078 \textit{k}-points provide reliable convergence across the full spectral range for this material, particularly below 4 eV. Minor oscillations remain at higher energies, but these do not compromise the overall accuracy.

Convergence with respect to the number of neighbors is analyzed in figure~\ref{fig:convergence}(c), where we use a \verb|DZP| basis, 1078 \textit{k}-points, and include 4, 16, 28, 34, and 46 neighbors (\textsf{neigh}=1, 2, 3, 4, or 5, respectively) in estimating the momentum matrix elements defined in equation~\eqref{eq:puv_modf}. The curves indicate that convergence is essentially reached slightly when considering up to second nearest neighbors.

From these analyses, a fully converged calculation for SiC can be obtained with a combination of a \verb|DZP| basis, 46 neighbors (\textsf{neigh}=5), and a 1078-point \textit{k}-grid (\textsf{nk}=30). This result is shown in figure~\ref{fig:convergence}(d), where solid lines are our results and the dotted lines are those from Ref.~\cite{Wu2008}. A key feature in this comparison is the prominent maximum of $\chiq$ just above the onset of $\mathrm{Im}[\chiq]$, which reflects the dominant frequency in the SHG response. This peak, labeled (II) in figure~\ref{fig:convergence}(d), appears near 3.5 eV. Its position results from a combination of band-structure effects--see figure~\ref{fig:bands}-—and the magnitude of the momentum matrix elements~\cite{HarrisonElcStr}. The transitions responsible for this feature primarily involve states near the $\Gamma$ point, where the density of states is enhanced due to a high concentration of nearby valence and conduction bands. Smaller structures in the spectrum, such as the shoulder at (I) and the dip at (III) in figure~\ref{fig:convergence}(d), similarly originate from transitions near the high-symmetry points $X$ and $L$ in figure~\ref{fig:bands}(a).

All curves in figure~\ref{fig:convergence} include both the $A_{ijk}$ and $B_{ijk}$ contributions. As shown in figure~\ref{fig:ABxyz}, the imaginary part of $A_{ijk}$ almost entirely governs the frequency dependence of $\chiq$ up to roughly 4.5 eV. This dominance arises because the $A_{ijk}$ term contains the Dirac delta function $\delta(\epsilon_{cv}-2\omega)$, making it particularly significant at lower frequencies.

To validate this approach further, $\chiq$ was calculated for gallium arsenide (GaAs), another zinc-blende material. Its lattice vector is $a = 5.62\,$\AA. Since the \verb|TZP| basis set was found not to offer a substantial improvement over \verb|DZP|, the latter was adopted. The GGA exchange--correlation functional and the Monkhorst--Pack $k-$point grid were kept identical to those used for 3C-SiC for the initial SIESTA calculation.

DFT significantly underestimates the band gap for GaAs---in our case yielding 0.6 eV. To recover the experimental value of 1.52 eV~\cite{Huang1992}, a (local) scissor correction was therefore applied to the conduction bands~\cite{Hughes1996}, shifting them by $\delta E$ according to $\e'_c = \e_c + \delta E$. For GaAs, a value of $\delta E = 0.91\,$eV reproduces the experimental gap. This procedure implicitly assumes that the conduction Bloch wavefunctions remain unchanged upon shifting, an approximation that is often acceptable when the gap correction mimics the effect of a GW calculation~\cite{Hybertsen1986}. A more rigorous treatment would require including the anomalous velocity induced by the scissor operator~\cite{Nastos2005, Cabellos2009} and is beyond the purpose of this work.

\begin{figure}[tb]
    \centering
    \includegraphics[width=0.95\linewidth]{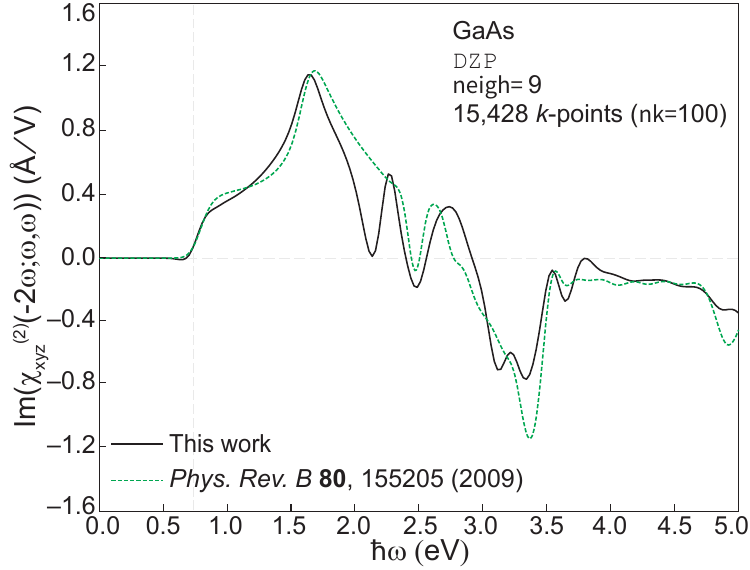}
    \caption{$\text{Im}[\chi^{(2)}_{xyz}]$ for GaAs as a function of the photon energy $\hw$. The dark curve corresponds to this work, while the dashed, lighter curve shows the result from Ref.~\cite{Cabellos2009}.}
    \label{fig:GaAS_xyz}
\end{figure}

Figure~\ref{fig:GaAS_xyz} shows the imaginary part of $\chi^{\scp (2)}_{xyz}$ obtained with our method (solid line), together with results from Ref.~\cite{Cabellos2009} (dashed line). As noted by several authors~\cite{Huang1993}, GaAs poses a more stringent challenge due to the structure of equations~\eqref{eq:shg1} and~\eqref{eq:shg2}, requiring a particularly fine sampling of the trIFBZ. For this reason, the calculation in figure~\ref{fig:GaAS_xyz} employs a denser mesh of 15428 $k$-points and neighbors up to a distance $a_9 = 7.94\,$\AA{} (\textsf{neigh}=9); the remaining parameters are \textsf{w} = 0.2 eV and $N = 1$ in equation~\eqref{eq:Dirac}.

Up to approximately 2 eV, our results agree well with previous calculations employing localized orbitals~\cite{Dumitrica1998,Huang1992} and plane waves~\cite{Cabellos2009,Hughes1996}. Above this energy, oscillations and sharp features appear, though they tend to average to the profiles reported in the literature. These structures originate from the phenomenon of {\it double resonances}~\cite{Nastos2005}, whereby certain regions in $k$-space simultaneously satisfy $\e^0_{cv} = \hw$ and cause denominators in equations~\eqref{eq:shg1} and~\eqref{eq:shg2} to vanish. To handle these divergences, a small imaginary part was added to the denominators of these terms. The prefactor $1/(\e^0_{cv})^3$ does not pose difficulties, as its denominator remains well above the band gap and can effectively be removed from the integral due to the accompanying Dirac delta. After taking the real part, this regularization is equivalent to
\begin{equation}
    \frac{1}{2\e^0_{cv'}-\e^0_{cv}}
    \;\rightarrow\;
    \frac{2\e^0_{cv'}-\e^0_{cv}}{(2\e^0_{cv'}-\e^0_{cv})^2+\eta^2},
\end{equation}
in the first term of equation~\eqref{eq:shg1}, with analogous modifications applied to the remaining three resonant denominators in equations~\eqref{eq:shg1} and~\eqref{eq:shg2}. The results in figure~\ref{fig:GaAS_xyz} were obtained with $\eta = 15\,$meV and show the characteristic peak$-$dip structure typically observed in zinc-blende materials~\cite{Hughes1996,Huang1993}. Finally, the local scissor correction employed here does not alter the Bloch wavefunctions and would therefore be expected to introduce a vertical scaling mismatch. However, this deviation is not palpably evident in figure~\ref{fig:GaAS_xyz}.

\subsection{Linear susceptibility}\label{sec:chi_linear}
As an extra validation of formalism described herein, and although not the central topic of this work, the linear susceptibility $\chil$ will be presented. Symmetry arguments similar to those leading to relations in equation~\eqref{eq:nonzero_chiq}, indicate that for a zinc-blende structure with $T_d$ point group, the only nonzero components of $\chil$ are~\cite{Boyd}:
\begin{equation}
    \chi^{\scp (1)}_{xx}=\chi^{\scp (1)}_{yy}=\chi^{\scp (1)}_{zz}.
\end{equation}

\begin{figure}[tb]
    \centering
    \includegraphics[width=\linewidth]{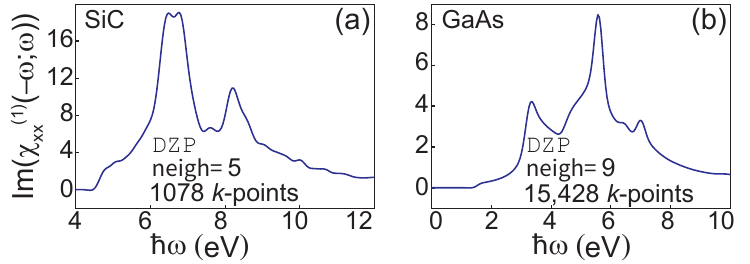}
    \caption{Imaginary part of the linear susceptibility $\chi^{\scp (1)}_{xx}$ as a function of the energy for (a) 3C-SiC and (b) GaAs.}
    \label{fig:chi_xx}
\end{figure}

The expressions for $\chi^{\scriptscriptstyle (1)}_{ij}$ are provided in Appendix~\ref{chi-derivation} (equation~\eqref{eq:chi_ij}).
Figure~\ref{fig:chi_xx} shows $\chi^{\scp (1)}_{xx}$ as a function of the energy $\hw$ for 3C-SiC and GaAs. In the last case, the (local) scissor operator to lift the conduction energy bands remains. The result for GaAs in figure~\ref{fig:chi_xx}(b) matches very well other results based on plane-waves~\cite{Hughes1996}. This is indicative that the momentum matrix elements are properly calculated, and that the erratic behavior of $\text{Im}[\xyz]$ in GaAs beyond 2 eV is indeed related to the almost zero denominators in equations~\eqref{eq:shg1} and~\eqref{eq:shg2}.

\subsection{Execution times}
We next provide runtime information, using 3C-SiC as an example. The times discussed below pertain to the calculation of the energy bands and momentum matrix elements all over the $k-$points sampling the trIFBZ, as indicated in the block MATRIX ELEMENTS AND LINEAR AND NONLINEAR OPTICAL RESPONSES in figure~\ref{fig:flowchart}; the most time-consuming part of the whole calculation.

\begin{figure}[t!]
	\centering
	\includegraphics[width=\columnwidth]{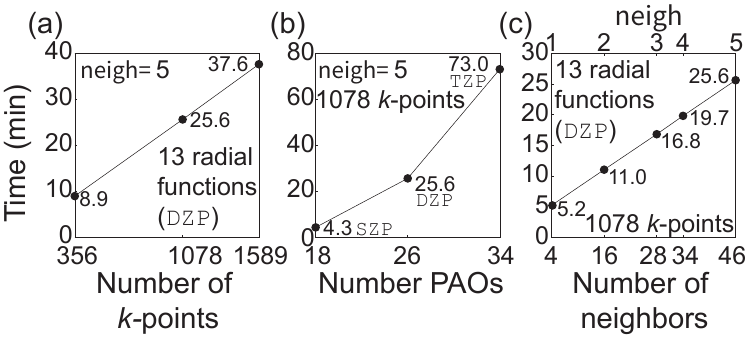}
	\caption{Execution times as a function of (a) the number of integration $k-$points in the trIFBZ, (b) the PAO basis size, and (c) the number of neighbors in the calculation of the momentum matrix elements (equation~\eqref{puv}).}
		\label{fig:times}
\end{figure}

Execution times in figure~\ref{fig:times} are examined as a function of the number of $k$-points, the basis size (or equivalently, the number of radial functions per atomic species), and the number of neighbors entering $\Pi_{j;ab}^j(\bm{k})$. The times in cases (a) and (c) grow  when increasing every one of these parameters in a close-to-linear manner.

\section{Conclusions}\label{conclusions}
A numerical approach for the calculation of the second-order optical susceptibility reliant on the symmetries of linear momentum integrals among PAOs was developed and demonstrated here. Appropriate Bloch eigenstates and matrix elements were obtained by solving a generalized eigenvalue problem with parameters from a DFT calculation. Two-center integrals ({\em i.e.}, matrix elements of the momentum operator among PAOs, were calculated by solving a symmetry-informed minimum set of two-center integrals for orbitals separated along the $\hat{z}-$direction, and then rotating the axes into the bond direction. The optical susceptibility was integrated using a Gaussian-Hermite representation of the Dirac delta. The formalism was applied to cubic SiC and GaAs, for which linear responses were reported as well. The results reproduce the susceptibility found in the literature.

\section{Acknowledgments}\label{acknowledgments}
This work was supported by the Department of Energy (grant DE-SC0022120). We acknowledge  conversations with Professors Bernardo Mendoza and Jos{\'e} Luis Cabellos. We also acknowledge multiple rounds of feedback and access to unpublished data from Professor Taisuke Ozaki, and the hospitality of the Institute of Solid State Physics at the University of Tokyo. Calculations were performed at the Arkansas High Performance Computing Center, funded by the NSF under Award OAC-2346752.

\appendix
\section{Perturbation expansion for the susceptibility}\label{chi-derivation}
Equations~\eqref{eq:shg1} and~\eqref{eq:shg2} are general expressions for $\chiq$ that make no assumption about the nature of the basis (orthogonal or non-orthogonal) for the expansion of the Bloch wavefunction. Thus, they can be obtained from a Hamiltonian in the $\bm{r}-$representation $\hat{H}_0=\hat{\bm{p}}^2/2m_e+V(\bm{r})$, where $m_e$ is the electron's mass and $V(\bm{r})$ is the periodic potential of the ionic background. If the applied optical field is described by the vector potential $\bm{A}(t)$, Peierls substitution leads to the time-dependent Hamiltonian $\hat{H}(t)$:
\begin{equation}
	\hat{H}(t)=\frac{1}{2m_e}\left( \hat{\bm{p}}+e\bm{A}(t)\right)^2+\hat{V}(\bm{r})\simeq \hat{H}_0+\hat{H}_1(t),
\end{equation}
where $-e$ is the electron charge ($e>0$). The linear and second-order susceptibility tensors $\chi_{ijk}^{(2)}$ are defined, in S.I. units, in terms of the Fourier components of the induced electric polarization $P_i(t)=P_i(\w) e^{-\ci\tw t}+c.c.$ and incident electric field $E_i(t)=E_i(\w) e^{-\ci\tw t}+c.c.$, where $\tilde{\omega} = \omega+\ci\eta$ ($\eta>0$), as follows \cite{Boyd}:
\begin{equation}
P_i(\w) = \varepsilon_0 \left(\chi_{ij}^{(1)} E_j(\w) +\chi_{ijk}^{(2)} E_j(\w) E_k(\w)+\ldots \right),
\end{equation}
where $i,j,k$ represent the Cartesian coordinates. A sum on repeated indices is implied and $\varepsilon_0$ is the vacuum permittivity. When dealing with insulators, the macroscopic density current is associated with the induced polarization $\bm{P}$ by $\bm{J} = d\bm{P}/dt$~\cite{JacksonCE}, and its expectation value can be written as:
\begin{align}\label{eq:J}
\braket{\bm{J}} &= \frac{1}{\Omega} \text{Tr} (\hat{\rho} \bm{\hat{j}}),
\end{align}
where $\hat{\rho}$ and $\bm{\hat{j}}$ are the density and current density operator, respectively, and $\Omega$ is a normalization volume. Using time-dependent perturbation theory, both $\hat{\rho}$ and $\bm{\hat{j}}$ can be expanded as a series using the vector potential $\bm{A}(t)$ as a small parameter:
\begin{align}
    \hat{\rho}=\hat{\rho}^{(0)}+\hat{\rho}^{(1)}+\hat{\rho}^{(2)}+\cdots \notag\\
    \bm{\hat{j}}=\bm{\hat{j}}^{(0)}+\bm{\hat{j}}^{(1)}+\bm{\hat{j}}^{(1)}+\cdots
\end{align}
Expanding equation~\eqref{eq:J} up to second-order in $\hat{\rho}$ and $\bm{\hat{j}}$ one gets:
\begin{align}\
\braket{\bm{J}^{(1)}} &= \frac{1}{\Omega}\Tr (\hat{\rho}^{(0)} \bm{\hat{j}}^{(1)} )+ \frac{1}{\Omega} \Tr(\hat{\rho}^{(1)} \bm{\hat{j}}^{(0)}), \label{eq:j1}\\
\braket{\bm{J}^{(2)}} &= \frac{1}{\Omega}\Tr (\hat{\rho}^{(0)} \bm{\hat{j}}^{(2)} )+ \frac{1}{\Omega} \Tr(\hat{\rho}^{(1)} \bm{\hat{j}}^{(1)}) + \notag\\
    &\hspace{4cm}\frac{1}{\Omega} \Tr(\hat{\rho}^{(2)}\bm{\hat{j}}^{(0)}). \label{eq:j2}
\end{align}
These expectation values directly relate to the first and second-order induced polarization via $\bm{J} = d\bm{P}/dt$ and, to the first and second-order susceptibility tensors. To make this connection explicit, expressions for the density and current operators up to second-order are required, as they are shown next.

\subsection{First and second-order density operator}
Starting from the iterative solution of the von Neumann equation:
\begin{equation}\label{eq:vonNeumann}
   \hat{\rho}^{\prime(n)}(t) = \frac{1}{\textsf{i}\hbar} \int_{\infty}^{t} [\hat{H}_1^{\prime}(t^{\prime}),\hat{\rho}^{\prime(n-1)}(t^{\prime})] dt^{\prime},
\end{equation}
where primed operators denote quantities in the interaction picture, $\hat{O}^{\prime}(t) = e^{\textsf{i}\hat{H}_0t /\hbar} \hat{O}e^{-\textsf{i}\hat{H}_0t/\hbar}$, the matrix form of the first-order density operator can be expressed as:
\begin{align}
&\bra{\psi^0_{m\bm{k}}}\hat{\rho}^{\prime(1)}(t) \ket{\psi^0_{n\bm{k}}}  \equiv  \rho^{\prime(1)}_{mn}(\bm{k},t)= \nonumber \\
&\frac{1}{\textsf{i}\hbar} \int_{-\infty}^{t}   \bra{\psi^0_{m\bm{k}}} [\hat{H}_1^{\prime}(t^{\prime}),\hat{\rho}^{\prime(0)}(t^{\prime})] \ket{\psi^0_{n\bm{k}}} dt^{\prime},
\end{align}
where $\psi^0_{m\bm{k}}(\bm{r})$ is given in equation~\eqref{kwf2} and $\bra{\psi^0_{m\bm{k}}} \hat{O} \ket{\psi^0_{n\bm{k}}}  \equiv \int d\bm{r} [\psi^0_{m\bm{k}}(\bm{r})]^* \hat{O}\psi_{n\bm{k}}(\bm{r})$.

If the basis set were complete (which is, strictly speaking, not the case here), then an identity operator can be defined $\hat{I} = \sum_{\bm{k}} \sum_{l} \ket{\psi^0_{l\bm{k}}} \bra{\psi^0_{l\bm{k}}}$ into the commutator. The perturbation Hamiltonian is expressed as $\hat{H}_{1}^{\prime}(t^{\prime}) = \frac{e}{m_e} A_j(t^{\prime})  \hat{p}_j^{\prime}(t^{\prime})$, where $j$ denotes Cartesian components, and$-$as usual$-$ summation over repeated Cartesian indices is implied. In the frequency domain, the vector potential is expressed as $\bm{A}(t^{\prime}) = \bm{A}(\w) e^{-\textsf{i}\tilde{\omega}t^{\prime}}$. Using this representation, and substituting the definition of the density matrix together with the explicit interaction-picture forms of the operators, one obtains:
\begin{equation}\label{eq:rho1}
\rho_{mn}^{(1)}(\bm{k},t) =\frac{e}{m_e}   \frac{p_{j;mn}(\bm{k})f_{mn}}{\e^0_{mn}(\bm{k})-\tilde{\e}} A_j(\w) e^{-\textsf{i}\tilde{\omega}t},
\end{equation}
where $f_{nm} = f_n-f_m$ is the difference in occupation numbers between states $n$ and $m$, $\tilde{\e}=\hbar \tilde{\omega}$, $\e^0_{mn}(\bm{k})= \e^0_{m}(\bm{k})-\e^0_{n}(\bm{k})$ and $p_{j;mn}(\bm{k}) =\bra{\psi_{m\bm{k}}}\hat{p}_{j} \ket{\psi_{n\bm{k}}} $, $\e^0_n(\bm{k})$ is obtained by solving equation \eqref{eq:GEP}.
By an analogous, though more involved, procedure, the matrix representation of the second-order density operator can be obtained:
\begin{align}\label{eq:rho2}
&\rho_{mn}^{(2)}(\bm{k},t) = -\left( \frac{e}{m_e}\right)^2   \frac{A_i(\w)A_j(\w)}{\e^0_{mn}(\bm{k})-2\tilde{\e}} e^{-2\textsf{i}\tw t} \times \nonumber\\
& \left(\sum_{\substack{l=1 \\ (l\neq n)}}^{\textsf{nbands}} f_{ln}\frac{p_{i;ml}(\bm{k}) p_{j;ln}(\bm{k}) }{\e^0_{ln}(\bm{k})-\tilde{\e}} -\sum_{\substack{l=1 \\ (l\neq m)}}^{\textsf{nbands}} f_{ml} \frac{p_{j;ml}(\bm{k}) p_{i;ln}(\bm{k}) }{\e^0_{ml}(\bm{k})-\tilde{\e}}\right).
\end{align}

\subsection{Perturbation series solution of the current density operators}
Recalling the definition of the current density operator $\hat{\bm{j}} = -e \dot{\hat{\bm{r}}}$, the $i-$th component of $\hat{\bm{j}}$ is given by:
\begin{align}
e\dot{\hat{r}}_i&= \hat{j}_i(t)  = \frac{e}{\textsf{i} \hbar}[\hat{x}_i,\hat{H}] =  \frac{e}{\textsf{i}\hbar}[\hat{x}_i,\hat{H}_{0} + \hat{H}_{1}(t)] \\
& = \left(\frac{e}{m_e} \hat{p}_j+\frac{e^2}{m_e}A_j(t) \hat{I} \right) \delta_{ij}\nonumber\\
&= \frac{e}{m_e} \hat{p}_i+\frac{e^2}{m_e}  A_i(t) \hat{I}\nonumber,
\end{align}
where the relations $[\hat{x}_i, \hat{p}_j] = \textsf{i}\hbar \hat{I} \delta_{ij}$, $[\hat{x}_i, \hat{p}_j^{2}] = 2\textsf{i}\hbar \hat{p}_j \delta_{ij}$, and as $[\hat{x}_i,\hat{V}(\bm{r})] =0$ have been considered, and $\hat{I}$ is the identity operator.. This way:
\begin{equation}\label{eq:jj0}
\hat{j}_i^{(0)}(t) =  \frac{e}{m_e}   \hat{p}_i,
\end{equation}
and:
\begin{equation}\label{eq:jj1}
\hat{j}_i^{(1)}(t) = \frac{e^2}{m_e} A_i(t)\hat{I}.
\end{equation}

\subsection{Linear and second-order susceptibility tensors, $\chi^{(1)}_{ij}$ and $\chi^{(2)}_{ijk}$}
The derivation of the susceptibility tensors is lengthy, and only the essential steps needed to obtain the linear and second-order susceptibility tensors directly are outlined below.

Introducing equations \eqref{eq:rho1}, \eqref{eq:rho2}, \eqref{eq:jj0} and \eqref{eq:jj1} into equations \eqref{eq:j1} and \eqref{eq:j2}, one gets:
\begin{align}
\braket{J_i^{(1)}} &=  \Big(\frac{e^2}{ m_e^2} \int \dk \sum_{m=1}^{^{\textsf{nbands}}} \sum_{\substack{n=1 \\ (n\neq m)}}^{^{\textsf{nbands}}}  f_{nm} \frac{p_{i;nm}(\bm{k}) p_{j;mn}(\bm{k})}{\e^0_{mn}(\bm{k})-\tilde{\e}} \nonumber\\
&  -\frac{n_0 e^2}{m_e} \delta_{ij} \Big) A_j(\w) e^{-\textsf{i}\tilde{\omega}t},
\end{align}
and
\begin{align}\label{eq:J_i2}
\braket{J_i^{(2)}} &=  \frac{e^3}{ m_e^3} \int \dk \sum_{m=1}^{\textsf{nbands}} \sum_{\substack{n=1 \\ (n\neq m)}}^{\textsf{nbands}} \frac{p_{i;nm}(\bm{k})}{\e^0_{mn}(\bm{k})-2\tilde{\e}}\times \nonumber \\
& \Bigg(\sum_{\substack{l=1 \\ (l\neq n)}}^{\textsf{nbands}} f_{ln} \frac{p_{j;ml}(\bm{k}) p_{k;ln}(\bm{k})}{\e^0_{ln}(\bm{k})-\tilde{\e}}-\nonumber \\
&  \sum_{\substack{l=1 \\ (l\neq m)}}^{\textsf{nbands}} f_{ml} \frac{p_{k;ml}(\bm{k}) p_{j;ln}(\bm{k})}{\e^0_{ml}(\bm{k})-\tilde{\e}} \Bigg)A_j(\w)A_k(\w) e^{-2\textsf{i} \tw t}.
\end{align}
where the sums over $n$ and $m$ run over the full set of Bloch eigenstates, while that over $l$ (in equation~\eqref{eq:J_i2}) skips the terms $n$ and $m$ as indicated.  Given that $\braket{\bm{J}} = d\bm{P}(t)/dt$, then the first- and second-order susceptibility tensors become:
\begin{align}\label{eq:C1}
\chi^{(1)}_{i j }&= \frac{e^2\hbar^2}{\varepsilon_0 m_e^2 \tilde{\e}^2} \int \dk \sum_{m=1}^{\textsf{nbands}} \sum_{\substack{n=1 \\ (n\neq m)}}^{\textsf{nbands}} f_{nm}\frac{p_{i;nm}(\bm{k})p_{j;mn}(\bm{k})}{\e^0_{mn}(\bm{k})-\tilde{\e}} \nonumber \\
&-\frac{n_0\hbar^2 e^2}{m_e\varepsilon_0 \tilde{\e}^2}\delta_{ij},
\end{align}
and
\begin{align}\label{eq:C2}
\chi^{(2)}_{i j k}&= -\frac{\textsf{i} e^3\hbar^3}{2\epsilon_0 m_e^3 \tilde{\e}^3} \int \dk \sum_{m=1}^{\textsf{nbands}} \sum_{\substack{n=1 \\ (n\neq m)}}^{\textsf{nbands}} \frac{p_{i;nm}(\bm{k})}{\e^0_{mn}(\bm{k})-2\tilde{\e}}\times \nonumber \\
&\bigg(\sum_{\substack{l=1 \\ (l\neq n)}}^{\textsf{nbands}}  f_{ln} \frac{p_{j;ml}(\bm{k}) p_{k;ln}(\bm{k})}{\e^0_{ln}(\bm{k})-\tilde{\e}}-\nonumber \\
&\sum_{\substack{l=1 \\ (l\neq m)}}^{\textsf{nbands}} f_{ml} \frac{p_{k;ml}(\bm{k}) p_{j;ln}(\bm{k})}{\e^0_{ml}(\bm{k})-\tilde{\e}} \bigg),
\end{align}
where the integrals are over the FBZ. Note that these equations diverge when $\tilde{\e}\rightarrow 0$. This divergence is unphysical for semiconductors and can be removed as follows. For the linear susceptibility $\chil$, using a partial fraction decomposition of the form:
\begin{align}
\chi_{ij}^{(1)}=\frac{1}{\tilde{\e}^2} B_{ij}+\frac{1}{\tilde{\e}} C_{ij}+D_{ij},\nonumber
\end{align}
where:
\begin{align}
B_{ij} &=\frac{e^2\hbar^2}{\ve_0 m_e^2 }\int \dk\sum_{m=1}^{\textsf{nbands}} \sum_{\substack{n=1 \\ (n\neq m)}}^{\textsf{nbands}} f_{nm}\frac{p_{i;nm}(\bm{k})p_{j;mn}(\bm{k})}{\e^0_{mn}(\bm{k})}-\nonumber\\
&\frac{n_0\hbar^2 e^2}{\ve_0 m_e}\delta_{ij},\nonumber \\
C_{ij} &=\frac{e^2\hbar^2}{\ve_0 m_e^2 }\int \dk \sum_{m=1}^{\textsf{nbands}} \sum_{\substack{n=1 \\ (n\neq m)}}^{\textsf{nbands}} f_{nm}\frac{p_{i;nm}(\bm{k})p_{j;mn}(\bm{k})}{\e^0_{mn}(\bm{k})^2},\nonumber\\
D_{ij} &=\frac{e^2\hbar^2}{\ve_0 m_e^2 }\int \dk \sum_{m=1}^{\textsf{nbands}} \sum_{\substack{n=1 \\ (n\neq m)}}^{\textsf{nbands}} f_{nm}\frac{p_{i;nm}(\bm{k})p_{j;mn}(\bm{k})}{\e^0_{mn}(\bm{k})^2(\e^0_{mn}(\bm{k})-\tilde{\e})}.\nonumber
\end{align}
For materials with time reversal symmetry, it can be shown that $p^{*}_{i;lm}(\bm{k})=-p_{i;lm}(-\bm{k})$ and $\e^0_{mn}(\bm{k})=\e^0_{mn}(-\bm{k})$, so that $C_{ij}=0$ and that the coefficient $B_{ij}$ takes the form:
\begin{align}
B_{ij}=-\frac{e^2\hbar^2}{\ve_0}   \int \dk \sum_{n=1}^{\textsf{nbands}} f_n \left( \frac{1}{m_n^*(\bm{k})} \right)_{ij},\nonumber
\end{align}
where $\left( \frac{1}{m_n^*(\bm{k})} \right)_{ij}$ is the so-called effective mass tensor, defined as~\cite{KittelQTS}:
\begin{align}
    \left( \frac{1}{m_n^*(\bm{k})} \right)_{ij} &= \frac{\delta_{ij}}{m_e}-\frac{1}{m_e^2} \sum_{\substack{m \\ (m\neq n)}} \bigg (\frac{p_{i;nm}(\bm{k})p_{j;mn}(\bm{k})}{\e^0_{mn}(\bm{k})} \notag\\
    & +\frac{p_{i;mn}(\bm{k})p_{j;nm}(\bm{k})}{\e^0_{mn}(\bm{k})} \bigg)=\frac{1}{\hbar} \frac{\partial^2 \e_n^0(\bm{k})}{\partial k^i\partial k^j}.
\end{align}
For a semiconductor, $f_n$ is either 1 (valence) or 0 (conduction), then:
\begin{align*}
    B_{ij}&=-\frac{e^2\hbar^2}{\varepsilon_0 (2\pi)^3} \sum_n  \int_{\text{FBZ}} d\bm{k} \frac{\partial^2 \e_n^0(\bm{k})}{\partial k^i\partial k^j} \\
    &=-\frac{e^2\hbar^2}{\varepsilon_0 (2\pi)^3} \sum_n \int_{\partial {\text{FBZ}}} dS^i   \frac{\partial \e_n^0(\bm{k})}{\partial k^j},
\end{align*}
due to the periodicity of the band energies, $\e_n^0(\bm{k})$ and their derivatives cancel at opposite boundaries of the FBZ, yielding $B_{ij}=0$. Therefore, the first-order susceptibility tensor is given only by $D_{ij}$ by:
\begin{equation}\label{eq:chi1p}
\chi_{ij}^{(1)} = \lambda_1 \int \frac{d\bm{k}}{(2\pi)^3} \sum_{m=1}^{\textsf{nbands}} \sum_{\substack{n=1 \\ (n\neq m)}}^{\textsf{nbands}} f_{nm}\frac{p_{i;nm}(\bm{k})p_{j;mn}(\bm{k})}{\e^0_{mn}(\bm{k})^2(\e^0_{mn}(\bm{k})-\tilde{\e})},
\end{equation}
where $\lambda_1=\frac{e^2\hbar^2}{\ve_0 m_e^2}$. The $f_{nm}$ factor allows for the reduction to valence and conduction states, only. Using time-reversal symmetry and the property $1/(x\pm \textsf{i}\eta)=\mathcal{P}(1/x)\mp \textsf{i}\pi \delta(x)$, the imaginary part of equation~\eqref{eq:chi1p} can be written as:
\begin{align}\label{eq:chi_ij}
    \text{Im}[\chil]&=-\lambda_1 \pi \int \dk \times\notag \\
    &\sum_{v=1}^{\textsf{nvb}} \sum_{c=\textsf{nvb}+1}^{\textsf{nbands}}  \frac{\text{Re}[p_{i;vc}(\bm{k})p_{j;cv}(\bm{k})]}{\e^0_{cv}(\bm{k})^2} \delta(\e^0_{cv}(\bm{k})-\hw),
\end{align}
where $\hw>0$ is assumed, so that only the resonance $\e^0_{cv}(\bm{k})-\hw$ is kept. Coming back to equation~\eqref{eq:C2}, Cartesian indices $j, k$ can independently take the values of $x,y,z$. Then, swapping $i$ and $j$, the second-order susceptibility tensor satisfies $\chi^{(2)}_{i j k}= \chi^{(2)}_{i  k j}$; this is an {\em intrinsic permutation symmetry}. Then, by the symmetrization of Cartesian indices $j$ and $k$ as outlined below equation (\ref{eq:poperator}), one gets:
\begin{align}\label{eq:chi22}
\chi^{(2)}_{i j k}&=  -\frac{e^3\hbar^3}{\ve_0 m_e^3 } \int \frac{d\bm{k}}{(2\pi)^3} \sum_{m=1}^{\textsf{nbands}} \sum_{\substack{n=1 \\ (n\neq m)}}^{\textsf{nbands}} \frac{\textsf{i}}{2\tilde{\e}^3(\e^0_{mn}(\bm{k})-2\tilde{\e})}  \notag \\
&\times \left(\sum_{\substack{l=1 \\ (l\neq n)}}^{\textsf{nbands}} \frac{p_{i;nm}(\bm{k})  \{ p_{j;ml}(\bm{k}), p_{k;ln}(\bm{k}) \}  }{\e^0_{ln}(\bm{k}) -\tilde{\e}}  f_{ln}\right. \notag\\
&\left. -\sum_{\substack{l=1 \\ (l\neq m)}}^{\textsf{nbands}} \frac{p_{i;nm}(\bm{k})  \{ p_{j;ml}(\bm{k}), p_{k;ln}(\bm{k}) \}  }{\e^0_{ml}(\bm{k})-\tilde{\e}} f_{ml}\right),
\end{align}
to reduce the crowded notation, one defines $\lambda_2 =\frac{e^3\hbar^3}{ \ve_0 m_e^3}  $. Using time reversal symmetry, and splitting equation \eqref{eq:chi22} for positive and negative values of $k_x,k_y,k_z$ and using (explicit $\bm{k}$ dependence is omitted for brevity):
\begin{align}
    p_{i;nm}  \{ p_{j;ml}, p_{k;ln} \} &-p_{i;mn}  \{ p_{j;lm}, p_{k;nl} \}  =\notag\\ &2\textsf{i}\,\text{Im}\big( p_{i;nm}  \{ p_{j;ml}, p_{k;ln} \}\big),
\end{align}
one gets:
\begin{align}\label{eq:chi22div}
     &\chiqs=\lambda_2 \int \frac{d\bm{k}}{(2\pi)^3} \sum_{m=1}^{\textsf{nbands}} \sum_{\substack{n=1 \\ (n\neq m)}}^{\textsf{nbands}} \frac{1}{2 \tilde{\e}^3(\e^0_{mn}(\bm{k})-2\tilde{\e})} \times \notag\\
&\bigg( \sum_{\substack{l=1 \\ (l\neq n)}}^{\textsf{nbands}} \frac{ \text{Im}\big( p_{i;nm}(\bm{k})  \{ p_{j;ml}(\bm{k}), p_{k;ln}(\bm{k}) \}\big) }{\e^0_{ln}(\bm{k})-\tilde{\e}} f_{ln} \notag -\\
& \sum_{\substack{l=1 \\ (l\neq m)}}^{\textsf{nbands}} \frac{\text{Im}\big( p_{i;nm}(\bm{k})  \{ p_{j;ml}(\bm{k}), p_{k;ln}(\bm{k}) \}\big) }{\e^0_{ml}(\bm{k})-\tilde{\e}} f_{ml} \bigg).
\end{align}
Equation~\eqref{eq:chi22div} scales as $1/\tilde{\e}^3$, which diverges in the limit $\tilde{\e}\rightarrow 0$. To address this behavior, a partial fraction decomposition is applied:
\begin{align}
\frac{1}{\tilde{\e}^3(\e^0_{ln}(\bm{k})-\tilde{\e})(\e^0_{ml}(\bm{k})-\tilde{\e}) }=\frac{A}{\tilde{\e}} + \frac{B}{\tilde{\e}^2} + \frac{C}{\tilde{\e}^3} + \mathcal{F}(\tilde{\e}), \nonumber
\end{align}
where the coefficients $A$, $B$, $C$ and $\mathcal{F}(\tilde{\e})$ are determined. As demonstrated in reference~\cite{Ghahramani1991}, the terms proportional to $A$, $B$, and $C$ vanish, leaving only $\mathcal{F}(\tilde{\e})$ as a nonzero contribution. This leads to ($\bm{k}$ dependence omitted for brevity):
\begin{align}\label{eq:chi2p}
 &\chi^{(2)}_{i j k}=\lambda_2 \int \frac{d\bm{k}}{(2\pi)^3} \sum_{m=1}^{\textsf{nbands}} \sum_{\substack{n=1 \\ (n\neq m)}}^{\textsf{nbands}} \sum_{\substack{l=1 \\ (l\neq n)}}^{\textsf{nbands}} \times\notag\\
 &\text{Im}\left[ p_{i;nm}  \{ p_{j;ml}, p_{k;ln} \}\right]\left( \frac{D_1}{\e^0_{mn}(\bm{k})-2\tilde{\e}}  + \frac{E_1}{\e^0_{ln}(\bm{k})-\tilde{\e}} \right) f_{ln} \nonumber \\
&-\lambda_2 \int  \frac{d\bm{k}}{(2\pi)^3} \sum_{m=1}^{\textsf{nbands}} \sum_{\substack{n=1 \\ (n\neq m)}}^{\textsf{nbands}} \sum_{\substack{l=1 \\ (l\neq n)}}^{\textsf{nbands}} \text{Im}\left( p_{i;nm}  \{ p_{j;ml}, p_{k;ln} \}\right)\times \nonumber \\
&\left( \frac{D_2}{\e^0_{mn}(\bm{k})-2\tilde{\e}}  + \frac{E_2}{\e^0_{ml}(\bm{k})-\tilde{\e}} \right) f_{ml} ,
\end{align}
and:
\begin{align}
D_1 & = \frac{16}{\e_{mn}^{0^3}(\bm{k}) (2\e^0_{ml}(\bm{k})-\e^0_{mn}(\bm{k}))},\nonumber\\
E_1&=\frac{1}{\e_{ln}^{0^3}(\bm{k}) (\e^0_{mn(\bm{k})}-2\e^0_{ln}(\bm{k}))},\nonumber \\
D_2 & =\frac{16}{\e_{mn}^{0^3}(\bm{k}) (2\e^0_{ml}(\bm{k})-\e^0_{mn}(\bm{k}))},\nonumber\\
E_2&=\frac{1}{\e_{ml}^{0^3}(\bm{k}) (\e^0_{mn}(\bm{k})-2\e^0_{ml}(\bm{k}))}.\nonumber
\end{align}
By using the property $1/(x\pm \textsf{i}\eta)=\mathcal{P}(1/x)\mp \textsf{i}\pi \delta(x)$, the imaginary part of equation~\eqref{eq:chi2p} is:
\begin{align}
 &\text{Im}[\chi^{(2)}_{i j k}]= \lambda_2 \pi \int \frac{d\bm{k}}{(2\pi)^3} \sum_{m=1}^{\textsf{nbands}} \sum_{\substack{n=1 \\ (n\neq m)}}^{\textsf{nbands}} \sum_{\substack{l=1 \\ (l\neq n)}}^{\textsf{nbands}}f_{ln}\times \notag\\
&\text{Im}\left[ p_{i;nm}  \{ p_{j;ml}, p_{k;ln} \}\right] \left(  D_2 \delta(\e^0_{mn}-2\e) + E_2 \delta(\e^0_{ln}-\e) \right)  \nonumber \\
& - \lambda_2 \pi \int \frac{d\bm{k}}{(2\pi)^3} \sum_{m=1}^{\textsf{nbands}} \sum_{\substack{n=1 \\ (n\neq m)}}^{\textsf{nbands}} \sum_{\substack{l=1 \\ (l\neq n)}}^{\textsf{nbands}} \text{Im}\left[ p_{i;nm}  \{ p_{j;ml}, p_{k;ln} \}\right]\nonumber \\
&\times \left(  D_1 \delta(\e^0_{mn}-2\e) + E_1 \delta(\e^0_{ml}-\e) \right) f_{ml}, \nonumber
\end{align}
with $\e=\hbar \omega$. The factors $f_{ln}$ and $f_{ml}$ ensure that only interband (valence–conduction) transitions contribute, as in the linear case, then splitting the equation above into two different contributions, one resonant with $\e=\hw$, and the other with $2\e=2\hw$, one gets the equations~\eqref{eq:shg1} and~\eqref{eq:shg2} in the main text.

\section{Spherical harmonics and $D^{(j)}(\phi,\ti)$ matrices}\label{Dmatrices}
The spherical harmonics $Y_{\li\mi}(\bm{r})$ transform upon the rotation $P(\phi,\ti)$ [equation~\eqref{rotation}] according to $Y_{\li\mi}(P^T\bm{r})=\sum_{\mi'=-l}^\li D^{(1)}_{\mi'\mi}(\phi,\ti)Y_{\li\mi'}(\bm{r})$, where $D^{(\li)}(\phi,\ti)$ is a unitary matrix. For $\li=1$, $D^{(1)}(\phi,\ti)$ is given by:
\begin{equation}\label{Dl1}
	D^{(1)}(\phi,\theta)=\left(
	\begin{array}{ccc}
		e^{\ci\phi}\cos^2\frac{\ti}{2} & \frac{\sin\ti}{\sqrt{2}} &  e^{-\ci\phi}\sin^2\frac{\ti}{2} \\
		-e^{\ci\phi}\frac{\sin\ti}{\sqrt{2}} & \cos\ti & e^{-\ci\phi}\frac{\sin\ti}{\sqrt{2}}  \\
		e^{\ci\phi}\sin^2\frac{\ti}{2} & -\frac{\sin\ti}{\sqrt{2}} &  e^{-\ci\phi}\cos^2\frac{\ti}{2}
	\end{array}
	\right),
\end{equation}
If we use the real forms defined in equation~\eqref{eq:RealHarmonics} and in Table~\ref{ta:SphericalHarmonics}, we have:
\begin{equation}\label{rotationp}
	M_1(\phi,\ti)=\left(\begin{array}{ccc}
	\cos\phi         &  0       & -\sin\phi       \\
	-\sin\phi\sin\ti &  \cos\ti & -\cos\phi\sin\ti \\
	\sin\phi\cos\ti  &  \sin\ti &  \cos\phi\cos\ti
	\end{array} \right),
\end{equation}
with entries $M^l_{\;j;1}$. As it was noted before, this matrix is different from the rotation matrix because of the sign and basis convention of the sperical harmonics used in SIESTA.  For spherical harmonics with $\li=2$ we have:
\begin{widetext}
\begin{align}\label{Dl2}
		&D^{(2)}(\phi,\theta)=\notag\\
		&\left(
        	\begin{array}{ccccc}
			e^{2 \ci \phi} \cos^4 \frac{\theta}{2} & 2 e^{\ci \phi} \cos^3 \frac{\theta}{2} \sin \frac{\theta}{2} & \frac{1}{2} \sqrt{\frac{3}{2}} \sin^2 \theta & e^{-\ci \phi} \sin^2 \frac{\theta}{2} \sin \theta & e^{-2 \ci \phi} \sin^4 \frac{\theta}{2} \\
			-2 e^{2 \ci \phi} \cos^3 \frac{\theta}{2} \sin \frac{\theta}{2} & e^{\ci \phi} \cos^2 \frac{\theta}{2} (-1 + 2 \cos \theta) & \sqrt{\frac{3}{2}} \cos \theta \sin \theta & e^{-\ci \phi} (1 + 2 \cos \theta) \sin^2 \frac{\theta}{2} & e^{-2 \ci \phi} \sin^2 \frac{\theta}{2} \sin \theta \\
			\frac{1}{2} \sqrt{\frac{3}{2}}\, e^{2 \ci \phi} \sin^2 \theta & -\sqrt{\frac{3}{2}}\,e^{\ci \phi} \cos \theta \sin \theta & \frac{1}{4} (1 + 3 \cos 2\theta) & \sqrt{\frac{3}{2}}\,e^{-\ci \phi} \cos \theta \sin \theta & \frac{1}{2} \sqrt{\frac{3}{2}}\, e^{-2 \ci \phi} \sin^2 \theta \\
			- e^{2 \ci \phi} \sin^2 \frac{\theta}{2} \sin \theta & e^{\ci \phi} (1 + 2 \cos \theta) \sin^2 \frac{\theta}{2} & -\sqrt{\frac{3}{2}} \cos \theta \sin \theta & e^{-\ci \phi} \cos^2 \frac{\theta}{2} (-1 + 2 \cos \theta) & 2 e^{-2 \ci \phi} \cos^3 \frac{\theta}{2} \sin \frac{\theta}{2} \\
			e^{2 \ci \phi} \sin^4 \frac{\theta}{2} & - e^{\ci \phi} \sin^2 \frac{\theta}{2} \sin \theta & \frac{1}{2} \sqrt{\frac{3}{2}} \sin^2 \theta & -2 e^{-\ci \phi} \cos^3 \frac{\theta}{2} \sin \frac{\theta}{2} & e^{-2 \ci \phi} \cos^4 \frac{\theta}{2}
		\end{array}\right).
	\end{align}
\end{widetext}

\newpage
While the  real version of the matrix above is:
\begingroup
\setlength{\tabcolsep}{6pt} 
\renewcommand{\arraystretch}{1.2} 
\begin{widetext}
\begin{equation}\label{rotationd}
	M_2(\phi,\ti)=\left(
    \begin{array}{ccccc}
\cos\ti \cos2\phi & \cos\phi \sin\ti & 0 & -\sin\ti \sin\phi & -\cos\ti \sin2\phi \\
-\cos2\phi\sin\ti & \cos\ti\cos\phi & 0 & -\cos\ti \sin\phi & \sin\ti \sin2\phi \\
\frac{\sq}{2}\sin^2\ti\sin2\phi & -\frac{\sq}{2}\sin2\ti\sin\phi & \tfrac{1}{4}(1 + 3\cos2\ti) & -\frac{\sq}{2}\cos\phi\sin2\ti & \tfrac{1}{2}\sqrt{3}\cos2\phi\sin^2\ti \\
-\frac{1}{2}\sin2\ti\sin2\phi & \cos2\ti\sin\phi & \sqrt{3}\cos\ti\sin\ti & \cos2\ti\cos\phi & -\tfrac{1}{2}\cos2\phi\sin2\ti \\
\tfrac{1}{4}(3 + \cos2\ti)\sin2\phi & \frac{1}{2}\sin2\ti\sin\phi & \tfrac{1}{2}\sqrt{3}\sin^2\ti & \frac{1}{2}\cos\phi\sin2\ti & \tfrac{1}{4}(3 + \cos2\ti)\cos2\phi
\end{array} \right),
\end{equation}
\end{widetext}
\endgroup
with entries $M^l_{\;j;2}$. It must be noted that matrices~\eqref{Dl1} and~\eqref{Dl2} are the Hermitian conjugates (equivalently, the inverses) of the more commonly ones found in the literature~\cite{RoseETAM,TinkhamGTaQM}. The difference arises because in angular momentum theory it is more common to define the rotation of function $f(\bm{r})$ as $P(\phi,\ti)f(\bm{r})=f(P(\phi,\ti)\bm{r})$, instead of $P(\phi,\ti)f(\bm{r})=f(P^T(\phi,\ti)\bm{r})$. The latter definition is more suitable for the calculation of the two-center integrals in equation~\eqref{d_integrals}.

\section{Reducing integrals to the trIFBZ}\label{sec:reduc_BZ}
The reduction of the integrals over the FBZ, equations~\eqref{eq:shg1} and~\eqref{eq:shg2}, follows the procedure outlined in section~\ref{sec:FBZ_intg}. In this section we will give some details on its actual application to a zinc-blende material, $i.e$., a when the point group is $T_d$. This point group has 24 symmetry operations~\cite{LandauQMNRT}:
(i) the identity; (ii) $2\pi/3$ and $4\pi/3$ rotations around each of the following four directions (the four main diagonals of the cube in figure~\ref{fig1}(a)):
\begin{align}
  &(\bx+\by+\bz)/\sqrt{3}, \\
  &(\bx+\by-\bz)/\sqrt{3}, \\
  &(\bx-\by+\bz)/\sqrt{3}, \\
  &(\bx-\by-\bz)/\sqrt{3},
\end{align}
(iii) six mirror planes normal to the following directions (planes bisecting the cube in figure~\ref{fig1}(a) and parallel to a pair of its faces):
\begin{align}
    &(\bx+\by)/\sqrt{2},\;\; (\bx-\by)/\sqrt{2}, \\
    &(\bx+\bz)/\sqrt{2},\;\; (\bx-\bz)/\sqrt{2}, \\
    &(\by+\bz)/\sqrt{2},\;\; (\by-\bz)/\sqrt{2},
\end{align}
(iv) three $\pi$ rotations around the $\bx$, $\by$ and $\bz$ axes, and (v) $\pi/2$ and $3\pi/2$ roto-inversion around the  axes $\bx$, $\by$ and $\bz$. Each one of this operations has its usual matrix representation $M_s$ ($s=1,\cdots,24$), which will not be shown here.
Because of its simplicity, the linear susceptibility will be analyzed first. According to what it was said in section~\ref{sec:chi_linear}, only the $xx$ needs to be dealt with. The generic equation~\eqref{symmed} applied to the expression for $\chil$ in equation~\eqref{eq:chi_ij} results in:
\begin{align}\label{eq:chi_ij_red}
    &\text{Im}[\chi^{\scp (1)}_{xx}]=-2\lambda_1 \pi\hbar^2 \int_\text{trIFBZ} \dk \times\notag \\
    &\sum_{v,c} \sum_{s=1}^{24}\frac{\text{Re}[\Pi_{1;vc}(M_s\bm{k})\Pi_{1;cv}(M_s\bm{k})]}{\e^0_{cv}(\bm{k})^2} \delta(\e^0_{cv}(\bm{k})-\hw),
\end{align}
where for convenience we used numeric indices in the momentum matrix elements $\Pi_{i;vc}$. Also, in the energies entering equation~\eqref{eq:chi_ij_red} the symmetry property $\e_n(\bm{k})=\e_n(M_s\bm{k})$ has been used. Equation~\eqref{eq:p_rot} applied to the case above looks like this:
\begin{align}\label{eq:p_rot2}
	 \Pi_{1;vc}(M_s\bm{k})\Pi_{1;cv}(M_s\bm{k})=&M_{s,1l}M_{s,1r}\times\notag\\
     &\Pi_{l;vc}(\bm{k})\Pi_{r;vc}(\bm{k}).
\end{align}
Replacing equation~\eqref{eq:p_rot2} into equation~\eqref{eq:chi_ij_red}, one gets:
\begin{equation}\label{eq:final_chil}
    \text{Im}[\chi^{\scp (1)}_{xx}]=2\sum_{s=1}^{24} M_{s,1l}M_{s,1r}\text{Im}[\tilde{\chi}^{\scp (1)}_{lr}],
\end{equation}
where the notation $\text{Im}[\tilde{\chi}^{\scp (1)}_{lr}]$ stands for the same equation~\eqref{eq:chi_ij} but with an integration over the trIFBZ instead of the full FBZ. Using the matrices $M_s$, the sum in equation~\eqref{eq:final_chil} entails a lengthy and error-prone calculation, that is better done using, for instance, Mathematica. The final result is:
\begin{equation}\label{eq:final_chil2}
    \text{Im}[\chi^{\scp (1)}_{xx}]=16\big( \text{Im}[\tilde{\chi}^{\scp (1)}_{xx}]+\text{Im}[\tilde{\chi}^{\scp (1)}_{yy}]+\text{Im}[\tilde{\chi}^{\scp (1)}_{zz}]\big).
\end{equation}
As it can be seen, the high symmetry of the group $T_d$ leads to an important cancellation of most terms in equation~\eqref{eq:final_chil}. For the nonlinear susceptibility $\chiqs$ the procedure is entirely analogous.  In this case it holds that:
\begin{equation}\label{eq:final_chiq}
    \text{Im}[\chi^{\scp (2)}_{xyz}]=2\sum_{s=1}^{24} M_{s,1l}M_{s,2r}M_{s,3q}\text{Im}[\tilde{\chi}^{\scp (2)}_{lrq}],
\end{equation}
because $\chi^{\scp (2)}_{xyz}$ is the only nonzero independent component.  Now we have terms of the kind:
\begin{align}\label{eq:p_rot3}
	 \Pi_{1;vc}(M_s\bm{k})&\Pi_{2;cv'}(M_s\bm{k})\Pi_{3;v'v}(M_s\bm{k})= \notag\\
  &M_{s,1l} M_{s,2r} M_{s,3q}  \Pi_{l;vc}(\bm{k})\Pi_{r;cv'}(\bm{k})\Pi_{q;v'v}(\bm{k}).
\end{align}
After carrying out the sum in equation~\eqref{eq:final_chiq} using equation~\eqref{eq:p_rot3} we get the final result:
\begin{equation}\label{eq:final_chiq2}
    \text{Im}[\chi^{\scp (2)}_{xyz}]=16\big( \text{Im}[\tilde{\chi}^{\scp (2)}_{xyz}]+\text{Im}[\tilde{\chi}^{\scp (2)}_{yzx}]+\text{Im}[\tilde{\chi}^{\scp (2)}_{zxy}]\big),
\end{equation}
where the $\text{Im}[\tilde{\chi}^{\scp (2)}_{lrq}]$ are to be understood in the same ways as $\text{Im}[\tilde{\chi}^{\scp (1)}_{lr}]$.


%

\end{document}